\documentclass[preprint]{elsarticle}

\usepackage{lineno,hyperref}
\usepackage[utf8]{inputenc}
\usepackage{graphicx}
\usepackage{graphics}
\usepackage{float}
\usepackage{siunitx}
\usepackage{natbib}
\usepackage{blindtext}
\usepackage[nice]{nicefrac}
\usepackage[dvipsnames]{xcolor}
\usepackage{amsmath}
\usepackage{lscape}
\usepackage{csquotes}
\usepackage{placeins}
\usepackage{url}

\graphicspath{{./figures/}}

\modulolinenumbers[5]

\journal{arXiv}

\biboptions{authoryear}

\begin{document}

\begin{frontmatter}

\title{Experimental measurement of respiratory particles dispersed by wind instruments and analysis of the associated risk of infection transmission}

\author[1]{Oliver Schlenczek}

\author[1]{Birte Thiede}

\author[1]{Laura Turco}

\author[1,2]{Katja Stieger}

\author[3]{Jana M. Kosub}

\author[1,4]{Rudolf M\"uller}

\author[3]{Simone Scheithauer}

\author[1,2,5]{Eberhard Bodenschatz\corref{mycorrespondingauthor}}
\cortext[mycorrespondingauthor]{Corresponding author}
\ead{eberhard.bodenschatz@ds.mpg.de}

\author[1]{Gholamhossein Bagheri\corref{mycorrespondingauthor}}
\ead{gholamhossein.bagheri@ds.mpg.de}


\address[1]{Max Planck Institute for Dynamics and Self-Organization (MPIDS), G\"ottingen 37077 , Germany}
\address[2]{Institute for Dynamics of Complex Systems, University of G\"ottingen, G\"ottingen 37077 , Germany}
\address[3]{Institute of Infection Control and Infectious Diseases, University Medical Center, G\"ottingen 37075, Germany}
\address[4]{Institute for Music and Aerosols, Bad W\"unnenberg 33181, Germany}
\address[5]{Laboratory of Atomic and Solid State Physics and Sibley School of Mechanical and Aerospace Engineering, Cornell University, Ithaca, NY 14853, USA}

\begin{abstract}

Activities such as singing or playing a wind instrument release respiratory particles into the air that may contain pathogens and thus pose a risk for infection transmission. Here we report measurements of the size distribution, number, and volume concentration of exhaled particles from 31 healthy musicians playing 20 types of wind instruments using aerosol spectrometry and in-line holography in a strictly controlled cleanroom environment. We find that playing wind instruments carries a lower risk of airborne disease transmission than speaking or singing. We attribute this to the fact that the resonators of wind instruments act as filters for particles $>$\SI{10}{\micro\meter} in diameter. We have also measured the size-dependent filtering properties of different types of filters that can be used as instrument masks. Based on these measurements, we calculated the risk of airborne transmission of SARS-CoV-2 in different near- and far-field scenarios with and without masking and/or distancing. We conclude that in all cases where there is a possibility that the musician is infectious, the only safe measure to prevent airborne transmission of the disease is the use of well-fitting and well-filtering masks for the instrument and the susceptible person.
\end{abstract}


\end{frontmatter}


\section{Introduction}
Numerous contagious diseases are known to spread via exhaled infectious particles that are inhaled by a susceptible, e.g. influenza, tuberculosis, Severe Acute Respiratory Syndrome (SARS), and most recently the Coronavirus Disease 2019 (COVID-19), which is caused by the Severe Acute Respiratory Syndrome CoronaVirus 2 (SARS-CoV-2) \cite[][]{Leung2021, Morawska2020, Shankar2021}. Currently, the ongoing COVID-19 pandemic is still considered a public health emergency of international concern \cite[][]{WHO2022}. About 8\% of the persons tested positive for SARS-CoV-2 with no or very mild symptoms of COVID-19 (Alpha strain of SARS-CoV-2) had a very high viral load (more than $10^9$ genome copies of the SARS-CoV-2 virus per swab), which promotes disease transmission by highly-infectious asymptomatic individuals if high viral load is associated with high infectivity \cite[][]{Jones2021}. It is well-known that various respiratory maneuvers, like  speaking and singing produce particles \cite[see e.g.][and references therein]{Morawska2009, Pohlker2021, Bagheri2021database}. However, the fact that wind instruments also release respiratory particles \cite[see e.g.][]{Lai2011, HeEtAl2020, Kaehler2020, Parker2020, Spahn2021, Stockman2021, Abraham2021, McCarthy2021, Firle2021, Gantner2022} indicates that band or orchestra rehearsals or concerts are also associated with an enhanced risk of airborne infection transmission (hereafter called \enquote{infection risk}). Because of this potential risk, most forms of live music events were forbidden in most of 2020 in Germany and also other countries. Governmental rules forced opera houses and theatres to close, and instrument education / singing lessons could not take place as usual. This unfortunate situation is mainly due to the fact that much is unknown about the emitted amount of respiratory particles and effective measures to mitigate the risk of infection during live performances of singers and wind instrument players. 

Among the particle emission from different respiratory activities, not much attention was given to wind instrument players prior to the COVID-19 pandemic. It was found that playing a very simple brass instrument such as a vuvuzela releases much more particles than a person who is shouting \cite[][]{Lai2011}. A study by \citet{HeEtAl2020} has shown that many wind instruments release more particles during the play in comparison to normal breathing or speaking. Instruments in the low register (e.g. tuba, bassoon) typically had less particle emission compared to instruments in the high register (e.g. clarinet, trumpet). The selection of instruments examined by \citet{HeEtAl2020} covers approximately two orders of magnitude in particle emission from the lowest (tuba) to the highest emitter (trumpet). For wind instruments with a simple geometry, increasing loudness was associated with increasing particle emission, but the net effect of loudness on the emitted particle concentration was low \cite[][]{HeEtAl2020}. \citet{Abraham2021} also examined the pitch dependence and found negative correlation between pitch and particle emission for instruments in the low register (in particular the bass trombone and the bassoon) while a positive correlation was found between pitch and particle emission for instruments in the high register (clarinet, oboe, trumpet). Short notes played on the trumpet led to more particles compared to long notes of the same pitch. However, it is still unknown which notes or playing styles on which instrument lead to the highest emission of particles. A study by \citet{McCarthy2021} came to the conclusion that the particle size distributions from breathing and from playing a wind instruments are similar, and that the mean particle emission is also comparable. \citet{Stockman2021} however measured higher particle number concentration compared to the data shown in \cite{HeEtAl2020} for clarinets (2 times more), bassoon (4 times more), oboe (4 times more) and French horn (3 times more). The emitted particle number concentration from the trumpet measured by \cite{Stockman2021} was 3 times lower compared to \cite{HeEtAl2020}. Emitted particle number concentration shown in \cite{Firle2021} is also higher for playing flute or oboe compared to speaking. In all of the studies mentioned above, and also in \cite{Brosseau2022}, the conclusion was drawn that the highest number density of emitted particles from wind instruments is \textless \SI{1}{\micro\meter}. 

Furthermore, these studies \cite[except for][]{Firle2021} used an Aerodynamic Particle Sizer (APS) from TSI Inc. as the primary aerosol size spectrometer, which nominally measures the aerodynamic diameter of particles in a size range from 0.5 to \SI{20}{\micro\meter}. However, it cannot detect smaller than \SI{0.5}{\micro\meter} that may include a significant number of particles containing $\sim$\SI{0.1}{\micro\meter} SARS-CoV-2 virions and, as pointed out in \citet{Bagheri2021database} and other studies \cite[e.g. see][and references therein]{Pohlker2021}, APS suffers from low detection efficiency for larger \SI{4}{\micro\meter} particles. Furthermore, it is extremely important to perform measurements in a certified cleanroom with controlled relative humidity and temperature. For a quantitative assessment of the infection risk, either the size distribution of the dry residues or the size distribution of the particles right after emission  needs to be known as described in \cite{Nordsiek2020} since particles in any state in between exhalation and fully dried  would introduce huge uncertainties in the infection risk calculation. The importance of controlled measurements becomes clear when one considers that the particle volume decreases by a factor of about 90 (almost two orders of magnitude) from exhalation to fully dry \cite[][]{Bagheri2021database}. 

In terms of infection risk, the most frequently discussed scenario is a well-mixed room with accumulation of infectious particles.
The well-mixed room is recently used by several investigations \cite[e.g.][]{BUONANNO_2020,Nordsiek2020, Miller2020, Jimenez_COVID19app_CIRES_2020, heads, Firle2021, Kriegel2021}. \cite{Miller2020} as well as \cite{Kriegel2021} determined the emission rate of pathogen copies from retrospective analyses of superspreading events. This is an insightful simplification of the dose-response model, but it remains unclear which size fraction of emitted particles drives the release of pathogen copies into the room. To give a reasonable estimation on the efficacy of infection mitigation measures, the relationship between particle size and number concentration of pathogen copies needs to be known. The risk calculation done by \cite{Firle2021} was based on a constant conversion factor from pathogen copies to particle number concentration within a specified diameter range. Numerical simulations carried out by \cite{Stockman2021} have shown that the well-mixed room assumption is reasonable for indoor spaces of \SI{18}{\square\meter} area and \SI{63}{\cubic\meter} volume if the susceptible is not standing downwind of the infectious (i.e. the donor of pathogens) at close distance.

Apart from that, the near-field is of particular relevance in larger indoor environments \cite[see][and references therein]{Hejazi2021,Bagheri2021}, for which the the horizontal and vertical spread and dilution of particles released from the bell end of wind instruments becomes important. \cite{Spahn2021} investigated the airflow from wind instruments (contrabassoon, bassoon, bass clarinet, English horn, tenor saxophone, clarinet, alto flute, oboe, recorder, piccolo, tuba, French horn, trombone and trumpet) played by professional musicians and found that for all instruments measured the flow had relaxed to that of the room at a radius of more than \SI{1.5}{\meter} from the bell end. This question was also investigated by \cite{Abraham2021}, who found that the so-called aerosol influence zone (i.e. the distance up to which the released particles were detectable above the background in the Minnesota orchestra hall) was less than \SI{0.3}{\meter} for all instruments examined. These findings suggest that the source is highly localized before the emission plume gets mixed with the ambient air, mainly due to the wide opening of the instrument bells and the resulting low air speed.

As shown in the diagrams of measured wind speed and particle concentration in \cite{Abraham2021} and in the schlieren photos of \cite{Becher2020}, the approximation of the particle emission plume as a jet emanating from the bell end of a wind instrument seems reasonable. This is supported by \cite{Brosseau2022}  who with  Particle Image Velocimetry (PIV) found a jet flow geometry. At a distance between 2 to \SI{6}{\meter}, the center velocity of the air jet from the wind instruments reached room air draft level very much in agreement with the results by \cite{Spahn2021}. Particle dispersion to more than a meter away from the wind instrument was also reported by \cite{Gantner2022}. With this, the far-field is more reasonable for larger distances and well mixed room models otherwise turbulent dispersion and transports need to be considered \cite[][]{Hejazi2021}.

Due to low flow velocity and the tendency of aerosol-laden air from the wind instrument to move vertically upwards, the usage of air purifiers above the instruments seems to be very effective to mitigate accumulation of potentially infectious aerosol \cite[][]{Abraham2021}. In their computational fluid dynamics (CFD) simulations \cite{Abraham2021} treated each particle released from a wind instrument as infectious. The reduction of infectious particles by the usage of mobile air purifiers was also examined by \cite{NarayananYang2021} in a music classroom. Air purifiers have a higher cleaning efficacy at high concentration of aerosols, i.e. close to the source of the infectious particles \cite[][]{NarayananYang2021}. However, their disadvantage is emission of sound that may disturb the music played.

The use of a filter mask at the bell end of wind instruments was shown as a possible mitigation measure by \cite{Parker2020} and also examined and discussed by \cite{Abraham2021}, \cite{Stockman2021}, and \cite{Firle2021}. \cite{Parker2020}, \cite{Abraham2021} and \cite{Stockman2021} examined particle number concentration with and without a mask and report penetration for total number concentration.  Neither work provided information on size-specific penetration, so it is not possible to estimate the risk of infection \cite{Nordsiek2020, Bagheri2021}. \cite{Firle2021} found, in contrast to the results in \cite{Abraham2021} and \cite{Stockman2021}, that surgical masks on the instrument bell do not reduce the particle concentration emitted from flute, oboe, clarinet or trumpet. Again, no details on mask penetration versus particle size were given and mask fit to the instrument bell was not discussed. 

\bigskip
Our literature review shows that it is fair to say existing data are inconsistent when it comes to the concentration of particles produced by a given musical instrument in different studies, the importance of near-field versus far-field infection risk, and the influence of instrument masking on infection risk.  
In addition, it is critical to consider particle shrinkage in ambient air and its impact on risk of airborne disease transmission, as well as size-dependent filtration, which have not yet been considered.
The importance of particle shrinkage becomes clear when one considers that particle volume can decrease by a factor of 90 in a fraction of a second, e.g., from exhalation to measurement \cite[][]{Bagheri2021database}.
The impact of size-dependent filtration on infection risk has also been shown to be extremely important \cite{Bagheri2021}.
These problems can be attributed in part to due to the large variability in measurement methods, the different levels of environmental control, limitations of measurement instruments, uncontrolled sampling, and/or lack of representative statistics. 

Our primary goal in this study is to address these issues all at once. We have studied different types of wind instruments in a strictly controlled cleanroom and under controlled sampling conditions using aerosol spectrometry and in-line holography. 
We have studied the size and number concentration of particles produced in the air exhaled by 31 musicians playing a total of 20 wind instrument types. 
In total 300 minutes of spectrometry data and 24000 holograms to present the field-standardized, normalized number-size distribution in the size range of \SI{1.5}{} to \SI{40}{\micro\meter} diameter is collected. 
The particle shrinkage is taken into account by controlling the relative humidity of the sampling tubes and recalculating the wet diameter of the particles at the time of exhalation.
We characterized 12 types of filter pieces used as instrument masks and generated size-dependent filtration curves. 
Using these data, we performed a through analysis of the risk of transmission in the near field according to the upper bound concept presented in \citet{Bagheri2021} and also in the far field in a room for different masking and social distancing scenarios. 

In the following, we first describe the particle measurement setup, the experimental procedure, data processing, information about the musicians, and the infection risk model.
This is then followed by the Results and Discussion section, in which the particle size distributions measured from different wind instruments are presented in detail and compared with particle emission during normal breathing, speaking, and singing. Finally, the upper bound of infection risk is presented in different exposure scenarios with and without social distancing and/or instrument/susceptible masking in the near and far field. In contrast to  previous studies, we also estimate the the risk of infection for musicians playing the wind-instruments, as they are typically unable to protect themselves with face masks.  The manuscripts ends with a summary of the main findings and conclusions. 

\section{Materials and Methods}
\label{sec:methods}

\subsection{Experimental setup and measurement procedure}
\label{sec:experiments}

For measurement of the particulate matter emitted by playing wind-instruments, we use a similar setup as described in \cite{Bagheri2021database}. In contrast to previous studies \cite[e.g.][]{HeEtAl2020, McCarthy2021}, we measured the size distribution of the particle residues  (dry at RH  $<30\%$) and calculate the full wet size distribution at exit from the instrument from the diameter shrinkage factor derived in \cite{Bagheri2021database}, i.e. we multiply the dried diameter by a factor of 4.5 to obtain the diameter at the time of production (hereafter referred to as the exhaled diameter).  All measurements were performed in a nominal ISO Class 6 cleanroom (i.e. less than 1 million particles per \SI{}{\cubic\meter}). However, our measurements have shown that the cleanroom met at least ISO Class 4 criteria, i.e. less than 1000 \textgreater\SI{0.3}{\micro\meter} particles per \SI{}{\cubic\meter}.
\bigskip

\textbf{Setup} 

\begin{figure}[htbp]
	\centering
	\includegraphics[width=1\linewidth]{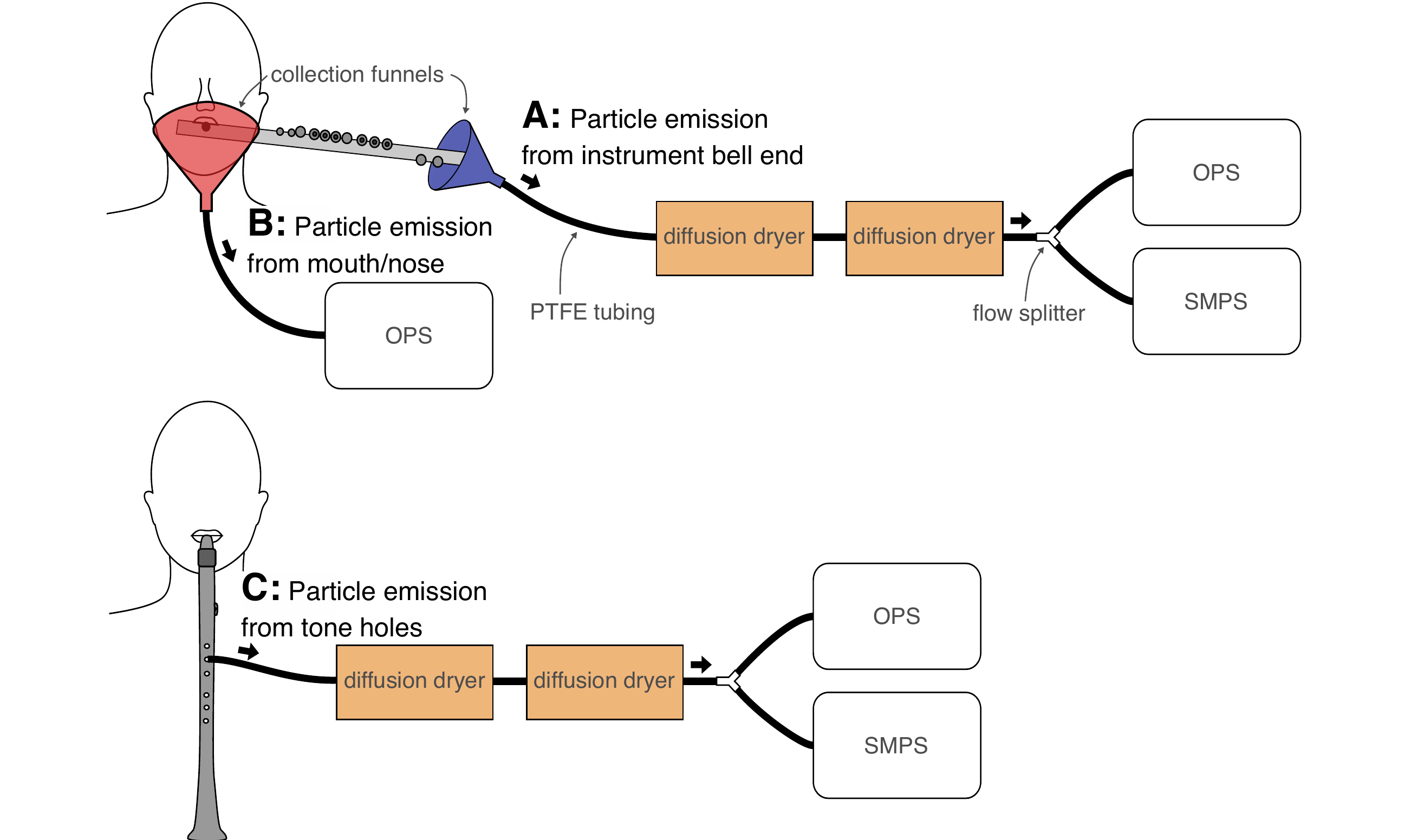}
	\caption{Schematics of measurement setups used inside the cleanroom for characterizing exhale particles produced by playing wind instruments. The standard configuration is setup A, where a funnel is placed at the bell end of the instrument and the particle-laden air passes two diffusion dryers (orange boxes) before reaching the aerosol size spectrometers. Setup B is used to measure particle emission from mouth and nose during the play of the wind instrument, here no dryer is used. Setup C is used for some woodwind instruments to measure leakage from the tone holes without a funnel connected to the tube end. Both setups B and C were optional while setup A was used for all instrument / subject combinations. If an instrument-mask was used, the funnel in setup A was placed onto the fabric (which covers the bell end) to avoid cleanroom air bypassing the funnel edge.}
	\label{fig:setup}
\end{figure}

 The main setup consisted of two aerosol size spectrometers, which were an Optical Particle Sizer Type 3330 (OPS) and a Nanoscale Scanning Mobility Particle Sizer Type 3910 (SMPS), both from TSI Inc. In the very beginning of the study, a third spectrometer, the  Aerodynamic Particle Sizer Model 3321 (APS) from TSI Inc., was also connected to the setup. The SMPS uses a flow rate of \SI{0.75}{\liter\per\minute} and the OPS uses \SI{1}{\liter\per\minute}. If the APS was connected to the setup, it added \SI{5}{\liter\per\minute} to the total flow rate (\SI{1}{\liter\per\minute} sample flow plus \SI{4}{\liter\per\minute} sheath flow). Both OPS and SMPS had sampling times of one minute per sample. While all experiments had valid OPS data, valid SMPS and APS data are only available for some of the measurements (see Table 1 in Supplementary Information (SI)). In the 16 logarithmically equidistant channels of the OPS, we obtain less resolution, but better counting efficiency (less losses) for particles larger than \SI{4}{\micro\meter} compared to an APS \cite[see][for more details]{Pohlker2021}, and we had two channels measuring particles smaller than \SI{0.5}{\micro\meter}. There is a reasonable agreement between OPS and APS up to \SI{4}{\micro\meter} particle diameter (see Fig. 2 of SI). According to the SMPS user manual, each of the 13 SMPS channels is scanned for approximately \SI{4.6}{\second} per sample. As the sample flow rate is \SI{0.75}{\liter\per\minute}, the sample volume for each channel in each one-minute sample is \SI{58}{\cubic\centi\meter}, which yields a physical detection limit of \SI{0.017}{\per\cubic\centi\meter} for the SMPS to measure one count. The spectrometers are connected via electrically conductive polytetrafluorethylene (PTFE) tubing with \SI{6}{\milli\meter} inner diameter to a flow splitter, which is connected via the same kind of tubing to two serial diffusion dryers (Grimm Model 7813), filled with silica gel. The outermost diffusion dryer is connected via the same kind of tubing to a sampling funnel made of plastic. Via intercomparison between the particle size distributions measured with the setup described in Fig. \ref{fig:setup} A and playing the instrument directly into the inlet nozzle of the OPS, we could verify that the material of the funnel (non-conductive plastic) did not introduce notable electrostatic losses (see Fig. 3 and 4 of SI). Particle losses due to the diffusion dryers are approximately 25\% and independent of the particle size (see \cite{Bagheri2021database} for more details). The total tube length was \SI{4.2}{\meter} from the funnel to each aerosol size spectrometer. For measuring the particle size distribution from wind instruments, we used two collection funnels of different size (a smaller one with \SI{100}{\milli\meter} diameter, and a larger one with \SI{150}{\milli\meter}), depending on the bell diameter of the instrument. We expect acceptable inlet efficiency around 50 - 80\% for the setup with the collection funnel for particles \textless \SI{3.5}{\micro\meter} (more details in Subsection 1.2. and Fig. 1 in SI). The standard setup for measuring the particle emission at the end piece or bell end of a wind instrument is shown in Fig. \ref{fig:setup}A. During those measurements where particle leakage from the nose or embouchure (i.e. next to the mouth) was measured in parallel with emission at bell end, a second funnel was placed in front of mouth and nose, connected to another OPS (see Fig. \ref{fig:setup}B).  For instruments with tone holes, we did additional measurements as described in Fig. \ref{fig:setup}C to determine the particle leakage from the tone holes in comparison to the particles emitted from the bell end. The role of leakage from mouth and nose, and the leakage from the tone holes is discussed in detail in Subsection 2.5. of the SI. Between the funnel and the bell end, there was a narrow gap for air to escape in order to operate the aerosol size spectrometers at ambient air pressure. As already studied by  \cite{Bouhuys1964}, the peak exhale volume of wind instrument players is much larger than the sampling flow rate used in our setup (up to \SI{100}{\liter\per\minute} for a tubist compared to a maximum sample flow rate of \SI{6.75}{\liter\per\minute}), so dilution due to the small gap at the side can be assumed to be negligible. In addition to the measurements of the particle emission at the open bell end, we examined the penetration of masks made of different materials similar to setup A in Fig. \ref{fig:setup}. Mask penetration was also measured with dolomite test dust as described in \cite{Bagheri2021} to obtain better counting statistics in particular for large particles. More details about the mask materials and their penetration can be found in Subsections 2.13. to 2.16. of the SI. In the following we call these filter pieces \enquote{instrument-masks}.

In addition to the dry particle measurements, the exhaled diameter of droplets within the first ten centimeters from the source was measured with the same holographic setup as described in \cite{Bagheri2021database}. The subject positions the bell end of the wind instrument or the tube end of a brass mouthpiece at about \SI{2}{} to \SI{3}{\centi\meter} upstream of the holographic sample volume and at the same horizontal level as the laser beam. The air with the particles crosses the sample volume of the holography unit and then approaches a collection funnel with an inner diameter of \SI{150}{\milli\meter}, which acts as aerosol sampling device and guides the particles that are small enough to follow the flow via the two dryers to the aerosol size spectrometers. Due to the larger distance between aerosol source and collection funnel, dilution will occur and these data are used only for qualitative comparison. Results of the holographic measurements can be found in Subsection 2.12. of the SI and in Subsection \ref{sec:origin}. The view volume of the holographic setup is about \SI{37}{\cubic\centi\meter} and its volume sampling rate is about \SI{13.6}{\liter\per\minute}. The holograms contain the diffraction pattern from each object in the sample volume, and via numerical reconstruction on a high performance computation (HPC) cluster, the size, location and shape of each particle from about \SI{6}{\micro\meter} to some \SI{}{\milli\meter} of particle diameter can be obtained \cite[details in][]{Fugal2009,Schlenczek2017}. With this setup, we  measured the particle size distribution and the local number density of particles $>$\SI{6}{\micro\meter}. Details about the size calibration of the holographic setup can be found in the Supplementary Information of \cite{Bagheri2021database}. 

\bigskip
\FloatBarrier
\textbf{Measurement procedure}

For each experimental run  the  subject(s) as well as the examiner(s) were required to wear ISO 4/5  clean-room overalls, hoods, and overboots. If possible they also wore clean-room gloves and face masks. All papers used were clean-room certified \cite[details in][]{Bagheri2021database}. Instruments. computers, and wind instruments were wiped with clean-room cloth to minimize contamination of the results by dust. Before the measurements in order to avoid artifacts from dust from inside the wind-instrument, they were either flushed with particle-free air, or were played for at least two minutes before the actual experiment started. In contrast to measuring particle emission from mouth and nose, for wind-instruments we need to account for the time of particle transport from the mouthpiece through the wind instrument to the particle spectrometers. Thus, the activities started at least \SI{30}{\second} before sampling was started. This time for particle transport was increased to \SI{90}{\second} for wind-instruments with large volume (e.g. B-flat tuba, see Fig. 6 in SI). Each measurement sample was \SI{60}{\second} long, and each experiment consisted of at least one sample. The subject played at least one piece of music selected by themselves, some also played single long notes at different pitch, or multiple pieces of music.

\subsection{Data processing}

For the OPS data, a detectability correction was applied. Basically, the sensitivity for \SI{0.3}{\micro\meter} and \SI{0.5}{\micro\meter} given in the calibration sheet was interpolated linearly to match the center diameter of the first two OPS channels. For \SI{0.3}{\micro\meter} particle diameter, the sensitivity was typically around 0.5, and for \SI{0.5}{\micro\meter} particle diameter, the sensitivity was close to 1. The sensitivity correction factor is the inverse of the sensitivity interpolated for the channel mid-points. With this sensitivity correction, a good agreement was found between optical diameter of the OPS and electrical mobility diameter of the SMPS (an example is shown in Subsection \ref{sec:origin}).

Due to the fact that the particle emission is concentrated in the center region of the holographic sample volume, the sample volume used for the data analysis has been restricted to $\pm$ \SI{3}{\centi\meter} in crosswise direction from the center of the emission plume (i.e. where most particles were found), and the outer \SI{2.5}{\milli\meter} on both sides in streamwise and vertical direction have been cut off. The remaining sample volume is $\SI{1.44}{\centi \meter} \times \SI{6}{\centi \meter} \times \SI{0.8}{\centi \meter} = $ \SI{6.9}{\cubic\centi\meter}. We also found that the sizes below \SI{8}{\micro\meter} have reduced detectability, so the concentrations below \SI{8}{\micro\meter} particle diameter in the holography data need to be interpreted with care. From \SI{8}{\micro\meter} to larger sizes, we did not see a notable reduction in detectability. In the size range above \SI{50}{\micro\meter}, we are typically limited by counting statistics when examining human exhale particles. 

\subsection{Subjects, instruments and music}

A call for participation was published via various communication channels. Flyers were displayed at the University Medical Center G\"ottingen (UMG), an announcement was posted on the UMG Facebook page, music students at the Hanover University of Music, Drama and Media (Hochschule f\"ur Musik, Theater und Medien Hannover HMTMH) were informed by their professors, and all staff involved in the study asked their friends to participate if they played a wind-instrument. An overview of the subjects, the instruments played and the number of experiments can be found in Table S1 of the SI.  A total of thirty-one musicians, ranging in age from 14 to 80 years, with an average age of 35 years, participated in the study. Five participants were from HMTMH, 8 played in a traditional German brass ensemble ("Blasorchester"), four were from a traditional German wind ensemble, four participants played in a trombone choir ("Posaunenchor"), 2 played in jazz bands and one played in a marching band ("Spielmannszug"). Due to fundamental differences in instrumentation and style between international wind ensembles, we decided to use the German term below in addition to the English. At the time the study was conducted, neither point-of-care tests nor vaccinations against SARS-CoV-2 were available. For all participants, an anamnesis was performed for possible risk situations due to close contact with persons infected with SARS-CoV-2 or travel to risk areas or virus variant areas as defined by the German Robert Koch Institute within the last 14 days, and they were asked about their health status.  During all phases of the study, the study team and the participants strictly adhered to the official regulations and the Institute's hygiene plan. With the exception of the musician whose particle emission was measured, all participants wore FFP2 masks throughout. None of the subjects felt ill on the day of the measurements and no COVID-19 infections were traceable to the study. Participants with symptoms or people closely exposed to someone tested positive for SARS-CoV-2 were excluded from the study. The study was ethically approved by the Max Planck Society. All subjects participated voluntarily, were informed in advance about the conduct of the experiment and subsequently consented to their participation. The subjects had to be legally competent and not impaired to carry out activities in this study. For participants below the age of 18 years, both the participants themselves and their legal guardians gave consent. Participation could be revoked at any time during the study.

\bigskip
\textbf{Instruments}

\begin{figure}[htbp]
	\centering
    	\includegraphics[width=0.9\textwidth]{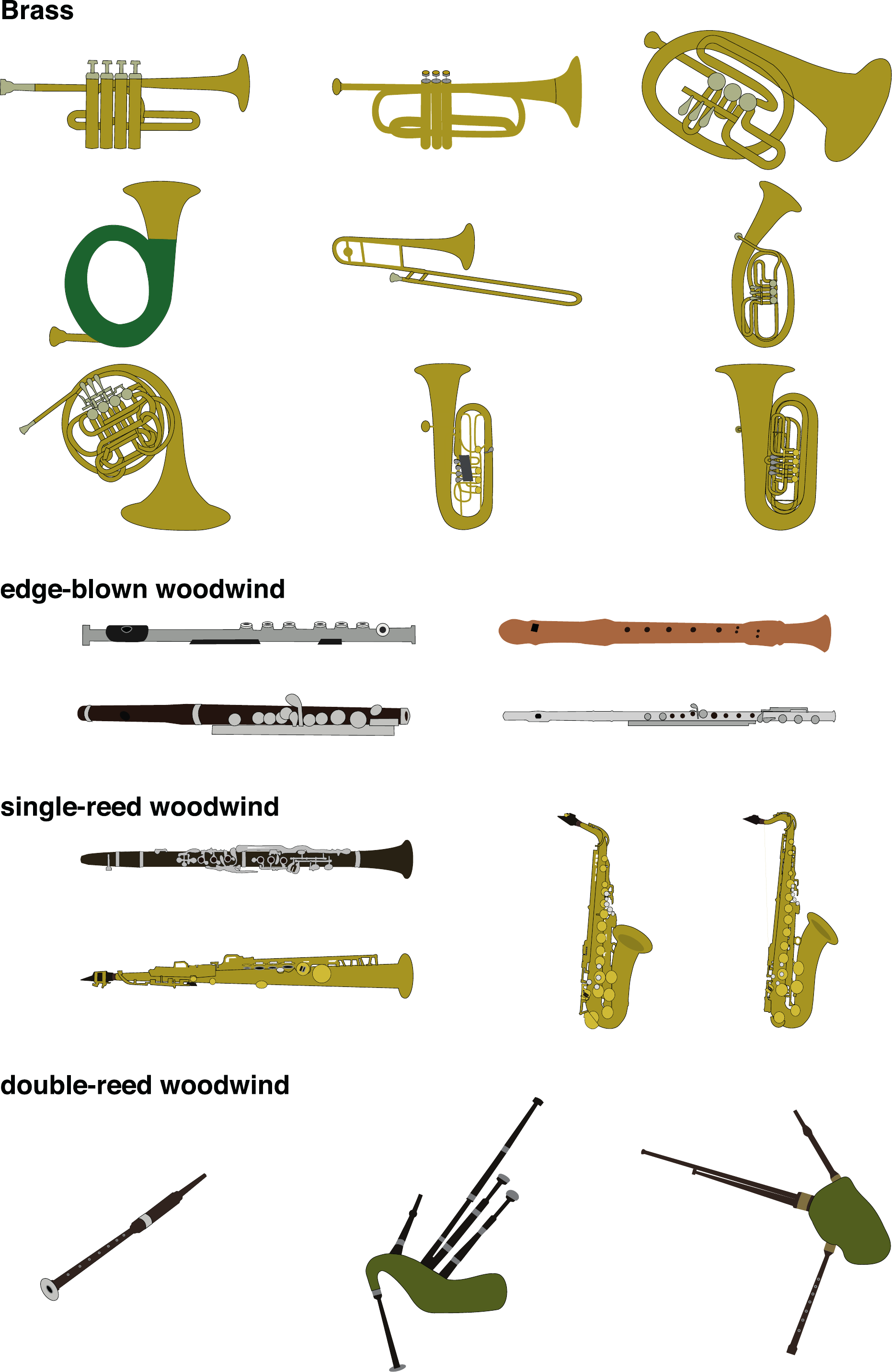}
	\caption{Illustration of the wind instrument types examined in this study grouped in different categories, namely brass instruments (top row: piccolo trumpet, trumpet, Kuhlo horn, middle row: hunting horn, tenor trombone, tenorhorn, bottom row: double horn, F tuba, Bb tuba), edge-blown woodwind instruments (top row: fife, alto recorder, bottom row: piccolo, flute), single-reed woodwind instruments (left: clarinet, soprano sax, middle: alto sax, right: tenor sax), and double-reed woodwind instruments (from left to right: practice chanter, Great Highland Bagpipes, Scottish Smallpipes).}
	\label{fig:instruments}
\end{figure}

For the purpose of this study we use the following categories for wind-instruments: \textit{Brass instruments} are wind-instruments where the tone is generated by fast opening and closing of the player's lips within the mouthpiece. Although the instruments are typically made of brass this also includes other materials like metals, plastics, or wood. \textit{Woodwind instruments} we put into three categories: \textit{edge-blown} , i.e.,  piccolo, flute, fife, and recorder ; \textit{single-reed}, i.e., clarinet and  saxophone; and  \textit{double-reed}, i.e.,  oboe, bassoon and  chanter of bagpipes. The sound from edge-blown woodwind instruments is created by blowing a fast air stream onto a sharp edge. In a single-reed instrument, the sound is created by the fast opening and closing of the air channel between the reed and the actual mouthpiece. Double-reed instruments use the same principle but the air channel is formed between both reeds. All clarinets in the study were clarinets in B-flat, all trumpets were trumpets in B-flat (with piston or rotary valves), and all trombones were tenor slide trombones in B-flat. The double horn in our study was played in B-flat only. The single-reed instruments in the study are all exposed single-reed instruments (the reed is actually in the player's mouth) while all double-reed instruments examined in this work are double-reeds in a windcap. An illustration of the wind instruments examined in our study is shown in Fig. \ref{fig:instruments}.

\bigskip
\textbf{Music}

The music pieces were chosen by the volunteers, each of whom chose something comparable to what they often play and have experience with. Some played pieces for symphony orchestras or small ensembles, others played pieces from traditional music, film music, jazz and military music. This differs from \cite{HeEtAl2020}, where a standardised pattern was played. In this study, we focused on music that is typically played over a long period of time, comparable to a rehearsal or concert situation for the volunteers. We chose this approach to better understand real-life  music events. This allowed us to identify scenarios with lower or higher emissions. Some subjects performed a series of experiments where they played long tones at different pitches to investigate the relationship between aerosol emission and the tone played (see Table S1 in the SI for more details).

\FloatBarrier
\subsection{Exposure Scenarios}
\label{sec:exposure}
Though there are four possible combinations of susceptible and infectious in a person-to-person exposure scenario, here we will mainly focus on an infectious musician (with \enquote{musician} we mean a wind instrument player) and a susceptible other person. For the different exposure scenarios, we need to differentiate between near-field and far-field exposure. The following subsections present in detail the methods and assumptions used to calculate the risk of infection from near- and far-field exposure in our study, but first we explain the general assumptions that apply to both exposure scenarios.

For the susceptible wearing a fitted FFP2 mask, we assume effective penetration as in \cite{Bagheri2021} for the \enquote{adjusted FFP2 mask}, which is fitted to the face without fit-test. From the different materials tested for filtering outflow of music instruments (see Subsection 2.13 with Figures S11 and S12 of SI), we found \enquote{filter05} to have the lowest penetration and decided to use this material for the instrument-mask. The effect of using other mask materials will be also discussed in the Results section. This material also had acceptable acoustic impedance according to the musicians who used it during the measurements. Size-specific penetration is taken into account by using the parameters obtained by the power-law fit for penetration of dry dolomite dust in Subsection 2.13. of SI. We assume that the infectious particles are already dry when they reach the mask material resulting in a lower filtration efficiency compared to the assumed diameter of the exhaled particles and thus a higher risk of infection. For all scenarios we assume that the wind instrument player plays the instrument without any interruption (i.e. no pause), which precludes a breakdown of the air jet emanating from the instrument bell and makes distancing less effective.
Effects from e.g. different breathing rates will be discussed in the respective Results subsection.

\subsubsection{Exposure Scenarios: near-field}

\begin{figure}[htbp]
	\centering
	\includegraphics[width=1.0\textwidth]{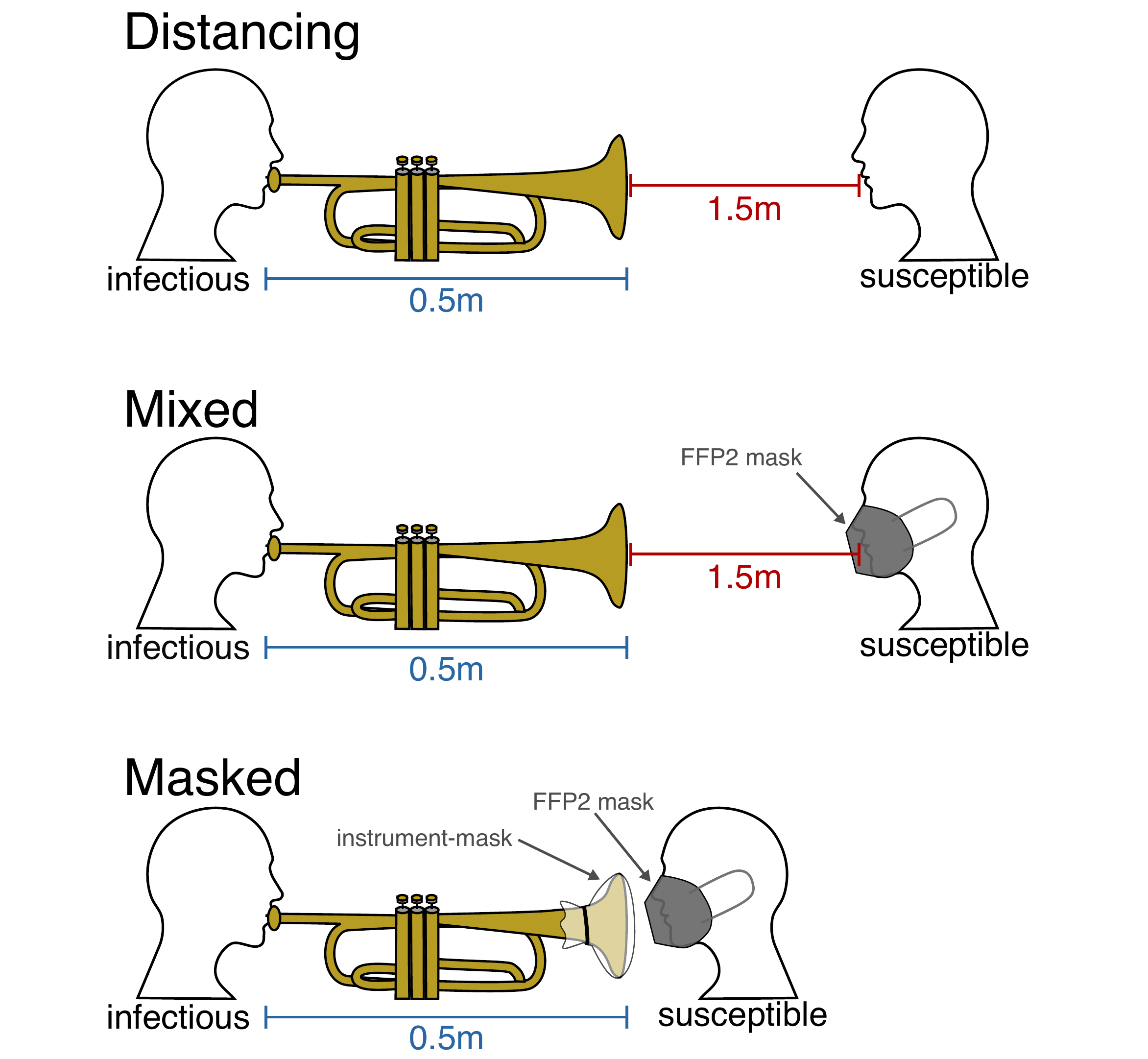}
	\caption{Exposure scenarios in the near field with one infectious musician (left) and one susceptible (right). Between the nose of the infectious and the instrument bell, a distance of $r_1 = $\SI{0.5}{\meter} (in blue) is assumed. In the distancing and mixed scenario, the susceptible keeps a distance of $r_2 = $\SI{1.5}{\meter} (in red) to the instrument bell. In the masked scenario, the wind instrument is covered with an instrument-mask, and the susceptible stands directly in front of the (covered) instrument bell, i.e. $r_1 = $\SI{0.5}{\meter}, $r_2 = $\SI{0}{\meter}. In the mixed and masked scenario, the susceptible wears an FFP2 mask fitted manually to the face without fit test as detailed in \cite{Bagheri2021}.} 
	\label{fig:scenarios}
\end{figure}

In all near-field scenarios, we assume that there is a distance of $r_1 = $ \SI{0.5}{\meter} between the face of the infectious musician and the bell end of the wind instrument. Also, we assume that there is an air-jet emanating from the nose during breathing, and from the bell end of the wind instrument during playing. It is assumed the musician is not speaking or singing. To give an upper bound on exposure, we assume that the air-jet from the nose goes horizontally towards the susceptible. In scenarios with distance, the distance between susceptible and instrument bell end will be $r_2 = $ \SI{1.5}{\meter}. We will consider three near-field exposure scenarios, which are illustrated in Fig. \ref{fig:scenarios}.

\begin{itemize}
    \item{\textbf{Distancing}: Instrument bell end of infectious and face of susceptible are separated by a distance $r_2 = $ \SI{1.5}{\meter} and oriented face-to-face. None of the two uses a mask.} 
    \item{\textbf{Mixed}: Instrument bell end of infectious and face of susceptible are separated by a distance $r_2 = $ \SI{1.5}{\meter} and oriented face-to-face. Susceptible is wearing an adjusted FFP2 mask.}
    \item{\textbf{Masked}: Instrument bell end of infectious and face of susceptible are separated by a distance $r_2 = $ \SI{0}{\meter} and oriented face-to-face. The instrument bell is covered with an instrument-mask and the susceptible is wearing an adjusted FFP2 mask. The air inhaled by the susceptible is primarily the undiluted air that was exhaled by the infectious through the instrument and filtered by the instrument-mask.}
\end{itemize}

For each scenario, we will show two extremes, which are a wind instrument with a high particle-volume-concentration emission and an instrument with a low particle-volume-concentration emission. They will be highlighted by \enquote{Max} and \enquote{Min}. Particle sizes from \SI{1.5}{} to \SI{45}{\micro\meter} exhaled diameter (\SI{0.3}{\micro\meter} to \SI{10}{\micro\meter} measured dry diameter ) are taken into account. 

\subsubsection{Exposure Scenarios: far-field}

\begin{figure}[htbp]
	\centering
    	\includegraphics[width=1.0\textwidth]{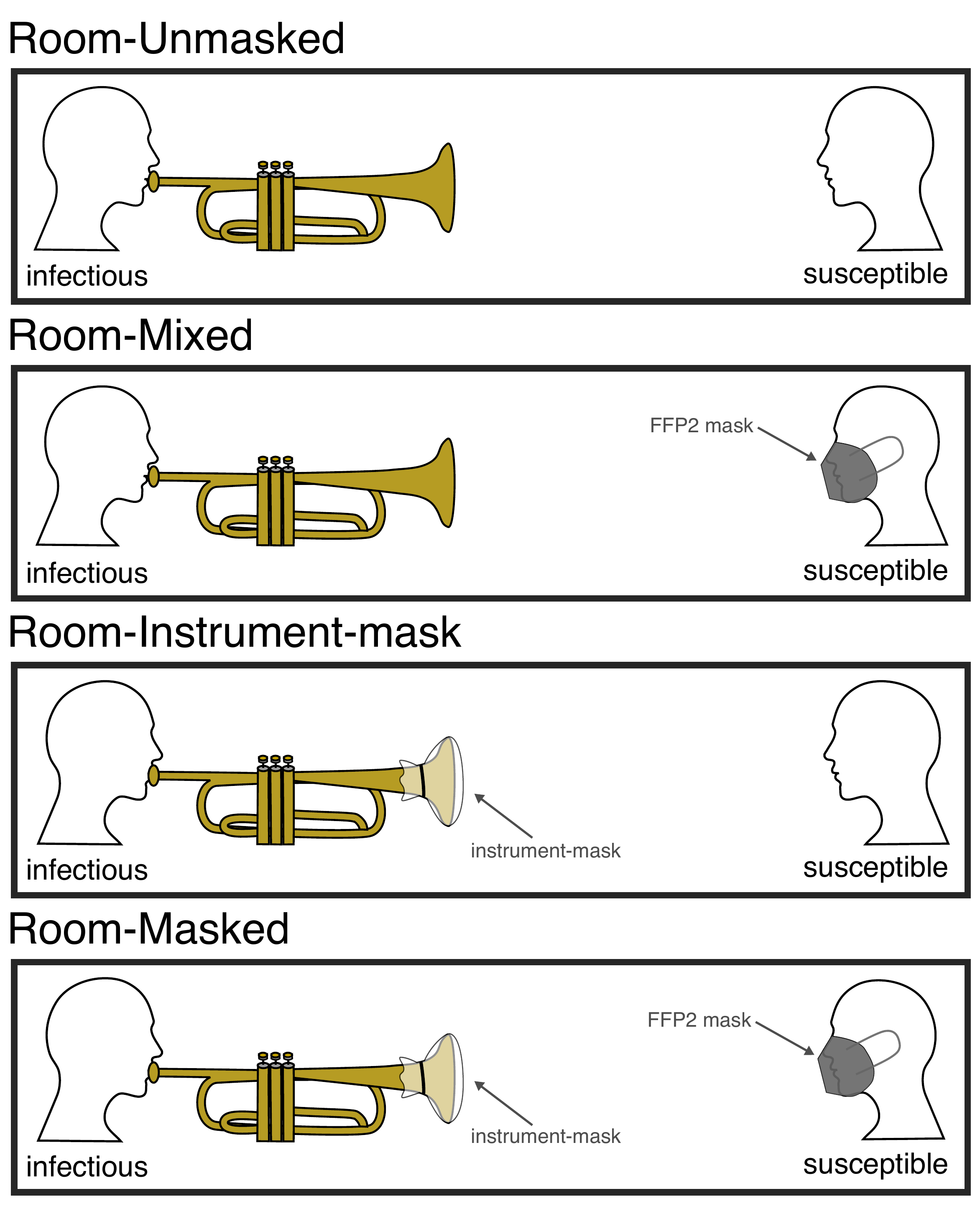}
	\caption{Illustration of the exposure scenarios in the far-field. Distances are assumed to be large enough such that no exposure to the near-field occurs. The scenarios are \enquote{Room-Unmasked} where neither infectious nor susceptible uses a mask, \enquote{Room-Mixed} where the susceptible wears a fitted FFP2 mask, \enquote{Room-Instrument-mask} where the infectious uses a mask for the instrument, and \enquote{Room-Masked} where both use masks. The room has a specific volume and also a specific ventilation rate.} 
	\label{fig:scenarios2}
\end{figure}

The far-field scenarios are not as complicated as the near-field scenarios because the distance between the infectious musician and the susceptible is by definition large enough to have no near-field exposure. 
This allows the use of well-mixed model, where it is assumed that all particles exhaled by the infectious musicians are immediately mixed throughout the whole volume of the room as long as the room is not too large, e.g., no larger than a classroom or small office. Here, we can examine all possible mask combinations as also shown in Fig. \ref{fig:scenarios2}:
\begin{itemize}
    \item{\textbf{Room-Unmasked}: Neither infectious musician nor susceptible use or wear a instrument/face mask.}
    \item{\textbf{Room-Instrument-mask}: Infectious musician uses an instrument-mask on the wind instrument, susceptible does not wear a mask.}
    \item{\textbf{Room-Mixed}: Infectious musician does not use an instrument-mask, susceptible wears an adjusted FFP2 mask.}
    \item{\textbf{Room-Masked}: Infectious musician uses an instrument-mask and susceptible wears an adjusted FFP2 mask.}    
\end{itemize}

\FloatBarrier
\subsection{Calculation of the risk of infection transmission}
\label{sec:infectionrisk}

By measuring the emitted particle size distribution from different musicians playing different wind instruments, we obtain the basis for calculating the risk of infection transmission (hereafter \enquote{infection risk}) of a susceptible person (i.e. a person who is exposed to a potential pathogen and can become infected) for SARS-CoV-2 in a typical live music environment if the musician is considered to be the infectious, i.e. the donor of pathogen copies. We assume that we have one susceptible sitting in the audience, and one infectious musician playing a wind instrument. It is assumed that the susceptible has not absorbed any SARS-CoV-2 copies prior to exposure. In this paper, we assume a dose-response relationship such that the absorption of a given pathogen dose by the susceptible will lead to some probability of infection. Even though individuals with some degree of immunity (e.g. vaccinated or recovered individuals) might not get infected at all after being exposed to a high number of pathogen copies, we assume that our susceptible individual does not have any immunity against the respective pathogen. This assumption is part of the upper bound concept as described in \cite{Bagheri2021}. As mentioned above, we consider both near- and far-field infection risk in a number of scenarios. As the risk of infection is the result of a stochastic process, adding more infectious or susceptible persons to the scenario increases complexity but does not significantly improve insight. Similar to the scenarios in \cite{Bagheri2021}, we examine the upper bound of infection risk for near-field exposure, but also for the far-field scenario. For the calculation of the risk of infection, the exponential model as described in \cite{Nordsiek2020} is used in a simplified way, i.e. the classical mono-pathogen model, as for SARS-CoV-2 the differences to consider the poly-pathogen are negligible. The mean infection risk $R_I$ is given by Eq. \ref{eq:1} where $\mu$ is the absorbed dose (which is time-dependent), and $\textnormal{ID}_{63.21}$ is the infectious dose required for 63.21\% chance of infection.
\begin{equation}
    R_I = 1 - \exp(- \mu / \textnormal{ID}_{63.21})
    \label{eq:1}
\end{equation}

 Here we assume $\textnormal{ID}_{63.21}=200$ as in \cite{Bagheri2021}, which is a good approximation of the infectious dose for the B.1.617.2 (Delta) strain of SARS-CoV-2. A discussion about $\textnormal{ID}_{63.21}$ and infection risk for different strains of SARS-CoV-2, e.g. wild type, Delta and Omicron variants, is given in Subsection 2.17 of SI. The absorbed dose $\mu$ depends on exposure duration $t$, inhaled volumetric flow rate of the susceptible $Q_{S}$ (which is the same as the exhaled volumetric flow rate), the concentration of pathogen copies in the inhaled air per particle diameter $n_{P,k}$, the dilution factor $f_d$ for mixing of the exhaled air with the ambient air before inhalation and the effective penetration $P_{I,k}$ of the mask the infectious is wearing, and the effective penetration $P_{S,k}$ of the mask the susceptible is wearing. Effective penetration means that total inward leakage is taken into account. Unmasked susceptible means $P_{S,k} = 1 \forall k$ and unmasked infectious means $P_{I,k} = 1 \forall k$. Also, the deposition within the respiratory tract $D_{rt,k}$ as a function of particle size needs to be known. Finally, $\mu$ can be expressed by Eq. \ref{eq:2} for $k = 1,2,...,K$ particle size channels considered. We assume conservation of mass in a sense that exhalation and inhalation flow rates are equal. 

\begin{equation}
    \mu = \underbrace{t}_{\text{exp. time}} \cdot \underbrace{Q_{S}}_{\text{inhal. susc.}} \cdot \underbrace{f_d}_{\text{dilution}} \cdot \sum_{k = 1}^{K}{\underbrace{n_{P,k}}_{\text{path. conc.}} \cdot \underbrace{P_{I,k}}_{\text{mask inf.}} \cdot \underbrace{P_{S,k}}_{\text{mask susc.}}\cdot \underbrace{D_{rt,k}}_{\text{dep. r.t.}}}
    \label{eq:2}
\end{equation}

 Values for $D_{rt,k}$ and $Q_{S}$ are taken from \cite{ICRP1994} while hygroscopic particle growth in the susceptible airways is also taken into account as explained in \cite{Bagheri2021} (to be discussed later). 
 
 To calculate $n_{P,k}$ for each size channel $k$, we need $N_k$ and also need to take into account that the particle diameter at exhale $D_0$ is by a factor of 4.5 larger than the particle diameter $D_{dry}$ measured after the diffusion dryers as described in \cite{Bagheri2021database}. We also need to know the concentration of pathogens $\rho_p$ in the airway surface liquid (ASL). For $D_{dry}$ in \SI{}{\micro\meter}, $\rho_p$ in pathogen copies \SI{}{\per\cubic\centi\meter} and $N_k$ in particles \SI{}{\per\cubic\centi\meter}, we get Eq. \ref{eq:4} with $n_{P,k}$ in the unit of pathogen copies per liter of exhaled air. 

\begin{equation}
    n_{P,k} = 10^{-9} \pi \rho_p N_k (4.5 D_{dry,k})^3 / 6
    \label{eq:4}
\end{equation}

Here, $N_k$ for the respective activity needs to be used (e.g. mean or maximum particle emission for playing a specific wind instrument). For $\rho_p$, we assume a value of $10^{8.5}$ \SI{}{\per\cubic\centi\meter}. There is some evidence that the viral load is not the same within the entire respiratory tract \cite[][]{Coleman2021}. Recent knowledge points at particles \textless \SI{5}{\micro\meter} dry diameter (up to \SI{22}{\micro\meter} diameter at exhalation) carrying the majority of virions exhaled for the Delta variant but particles originating from the lung are expected to have lower viral load for the Omicron variant \cite[see e.g.][]{chen2021omicron, Chan2021, Halfmann2022}. \cite{Coleman2021} found a high concentration of virions in the fine particle fraction for phonetic activities, but it remains unclear where the main source of infectious particles is located. It could for example be saliva, or the epithelial lining fluid (ELF) in the deep airways, or the mucus covering the vocal folds and the throat, or nasal discharge running from the nose into the throat.

 \subsubsection{Calculation of the infection risk in the near-field}
 \label{sec:nearfield}
 
 For a near-field exposure scenario, the dilution factor $f_d$ depends on the distance between infectious and susceptible. For a distancing scenario, $f_d$ can be calculated via Eq. \ref{eq:3} where $a$ is the aperture radius of the source (e.g. mouth radius or radius of bell end of a wind instrument), $r$ is the distance between source and susceptible, and $\alpha$ is the cone half-angle of the exhaled air jet as described in \cite{Yang2020}.
 
\begin{equation}
    f_d = a/(r \tan(\alpha))
    \label{eq:3}
\end{equation}

Here we chose $a = $\SI{1.8}{\centi\meter} and $\alpha =$ \SI{10}{\degree} for the emission from mouth or nose of the infectious \cite[same as in][]{Bagheri2021}, and $a = $\SI{3}{\centi\meter} and $\alpha =$ \SI{10}{\degree} for the emission from the wind instrument. This yields $f_d = 0.11$ for the emission from the wind instrument at $r = $ \SI{1.5}{\meter}, $f_d = 0.05$ for the emission from mouth and nose at $r = $ \SI{2}{\meter}, and $f_d = 0.2$ for the emission from mouth and nose at $r = $ \SI{0.5}{\meter}. If the infectious is using a mask, we assume undiluted exhaled air ($f_d = 1$) inhaled by the susceptible. a jet flow from the instrument bell and from mouth or nose of the infectious means implicitly that particle transport by convection is neglected. The role of convection is discussed for example in \cite{Hedworth2021} Also, the possibility of the susceptible being lucky as a non-zero background flow carries away the infectious particles is not considered here.

In terms of emission sources, we need to take into account that an infectious musician is not only playing the wind instrument (with infectious particles released mostly from the bell end) but also breathing, with infectious particles released from mouth and nose. By taking this into account, Eq. \ref{eq:2} needs to be modified. As the volumetric flow rate for exhalation through the wind instrument is typically different from the volumetric flow rate for normal breathing, we decided to take the volume-weighted average of emitted pathogen concentration at the location of the susceptible. The additional terms are the exhalation flow rate $Q_{w}$ of the infectious through the wind instrument, the exhalation flow rate $Q_{b}$ of the infectious breathing, the dilution factor $f_{d,w}$ for the exhale from the wind instrument, and $f_{d,b}$ for the exhale from breathing. Note that the specific emitted concentration of pathogen copies is $n_{P,k,w}$ for the wind instrument, and $n_{P,k,b}$ for breathing. Here, an instrument-mask is considered in the term $P_{I,k,w}$ and we assume that mouth and nose of the infectious are unmasked. Finally, $\mu$ for the near-field exposure can be calculated via Eq. \ref{eq:2b}.

\begin{equation}
    \mu = {t} {Q_{S}}\sum_{k = 1}^{K} (Q_{w} f_{d,w} {n_{P,k,w}} {P_{I,k,w}} + Q_{b} f_{d,b} n_{P,k,b}){P_{S,k}}{D_{rt,k}}/(Q_{w} + Q_{b})
    \label{eq:2b}
\end{equation}


\subsubsection{Calculation of the infection risk in the far-field}

A real-world exposure scenario is expected to be somewhere in between a near-field exposure and a far-field exposure scenario. We examine the infection risk in the far-field by assuming a small, well-mixed room. The room is assumed to have an area $A = $ \SI{12}{\square\meter} and a ceiling height of $h = $ \SI{2.5}{\meter}, which yields a volume $V = $ \SI{30}{\cubic\meter}. This room is assumed continuously cleaned from particles by an air purifier with an equivalent clean air delivery rate (CADR) of 4 air-changes-per-hour (ACH). Only the infectious and the susceptible are in this room. The initial concentration of pathogen copies in the room air is zero. In a well-mixed room of volume $V$, we use the pathogen concentration $n_{P,k}$ derived from the particle concentration $N_k$ and size to calculate the infection risk as explained in detail in \cite{Nordsiek2020}. The rate of change in $n_{P,k}$ as expressed in Eq. \ref{eq:5} is given by the source term (which is the emission of pathogen copies into the room by the infectious) and four sink term: (i) the removal of pathogens via inhalation by the infectious, (ii) the removal of pathogens via inhalation by the susceptible (this yields the absorbed dose per time for the susceptible, which determines the infection risk), (iii) the removal of pathogens due to ventilation of the room or usage of an air purifier with $\beta_v$ in the unit of air changes per second, and (iv) the pathogen inactivation $\gamma$, which is approximately \SI{0.64}{\per\hour} as described in \cite{vandoremalen}. A recent study by \cite{Oswin2022} shows that pathogen inactivation might depend on relative humidity and ambient CO$_2$ concentration and could actually be faster compared to the data in \cite{vandoremalen}. As we are interested in the upper bound on exposure to infectious particles, we use the more conservative inactivation rate by \cite{vandoremalen}.

\begin{equation}
    d n_{P,k} / dt = \underbrace{S_k}_{\text{emission}}  \underbrace{-Q_{I} D_{rt,k} n_{P,k}/V}_{\text{inhal. inf.}}  \underbrace{-Q_{S} D_{rt,k} n_{P,k}/V}_{\text{inhal. sus.}}  \underbrace{-\beta_v n_{P,k}}_{\text{ventilation}}  \underbrace{-\gamma n_{P,k}}_{\text{inactiv.}} 
    \label{eq:5}
\end{equation}

The reduction in pathogen concentration due to inhalation by the susceptible changes if the susceptible is wearing a mask. With $P_{S,k}$ being the effective mask penetration (including leakage), we get $-Q_S(1 - P_{S,k}^2 (1 - D_{rt,k}))n_{P,k}/V$ as the third term in Eq. \ref{eq:5}. 

For removal of infectious particles by an air purifier with a non-ideal filter, $\beta_v = Q_v (1 - P_{f,k}) / V$ with the clean air delivery rate $Q_v$, room volume $V$ and size-dependent filter penetration $P_{f,k}$. The source term $S_k$ may consist of multiple emission terms if there are combined activities, e.g. breathing and also playing a wind instrument. For both breathing and playing a wind instrument, we can write the source term as in Eq. \ref{eq:6}

\begin{equation}
    S_{k} = \underbrace{Q_{I,w} n_{P,k,w}}_{\text{path. from instrument}} \underbrace{P_{I,k}}_{\text{mask penetr.}} + \underbrace{Q_{I,b} n_{P,k,b}}_{\text{path. from breathing}}
    \label{eq:6}
\end{equation}

where $Q_{I}$ is the exhale volumetric flow rate with $w$ for playing wind instrument and $b$ for breathing, and the size-dependent emitted concentration of pathogen copies $n_{P,k,w}$ and $n_{P,k,b}$ for each activity. $P_{I,k}$ is the penetration of an instrument-mask which covers the instrument but not mouth and nose of the infectious musician. For an upper bound on exposure, we use $Q_{I,w}$ from \cite{Bouhuys1964}, and $Q_{I,b}$ from \cite{ICRP1994} where we assume relaxed breathing while seated ($Q_{I,b} = $ \SI{9}{\liter\per\minute}). This is the same volumetric flow rate as for the susceptible ($Q_S$) who breathes while seated.

The effect of ventilation or particle removal by an air purifier in this model is limited to the removal after the exhale has been mixed with the air of the entire room such that there is no concentration gradient anymore. In typical settings, however, the air cleaner or exhaust is located close to the source. If this is the case, the source term needs to be modified to allow direct capture of infectious particles, as shown in Eq. \ref{eq:7}. 

\begin{equation}
    S_{k} = (\underbrace{Q_{I,w} n_{P,k,w} P_{I,k}}_{instrument\; w/ \; mask} + \underbrace{Q_{I,b} n_{P,k,b}}_{breathing}) (1 - \underbrace{c}_{direct\;capture} \underbrace{(1 - P_{f,k})}_{air\; cleaner\; eff.})
    \label{eq:7}
\end{equation}

Here, $0 \leq c \leq 1$ is the fraction of particles captured directly, and $P_{f,k}$ is the size-dependent penetration of the filter which is used in the air cleaner. If a good filter is used ($P_{f,k}$ is small for all particle sizes), the effect of direct capture is effectively a linear reduction of the source strength, so the air cleaner would act like a mask. However, the flow pattern in the vicinity of an air cleaner is very complicated and it requires computational fluid dynamics (CFD) models or experiments under very controlled conditions to assess the capture efficiency $c$ as a function of distance between source and air cleaner, volumetric flow rate, and inlet and outlet geometry of the air cleaner. For a real exposure scenario, this would then still be limited to situations where few people are in the room and the airflow from the air cleaner is not altered by persons, furniture, air inlets or outlets, and other air cleaners.
\FloatBarrier
\section{Results and discussion}
\label{sec:results}

\subsection{Experimentally-measured size distribution of respiratory particles exhaled through wind instruments}
\begin{figure}[htbp]
	\centering
	\includegraphics[width=1.0\textwidth]{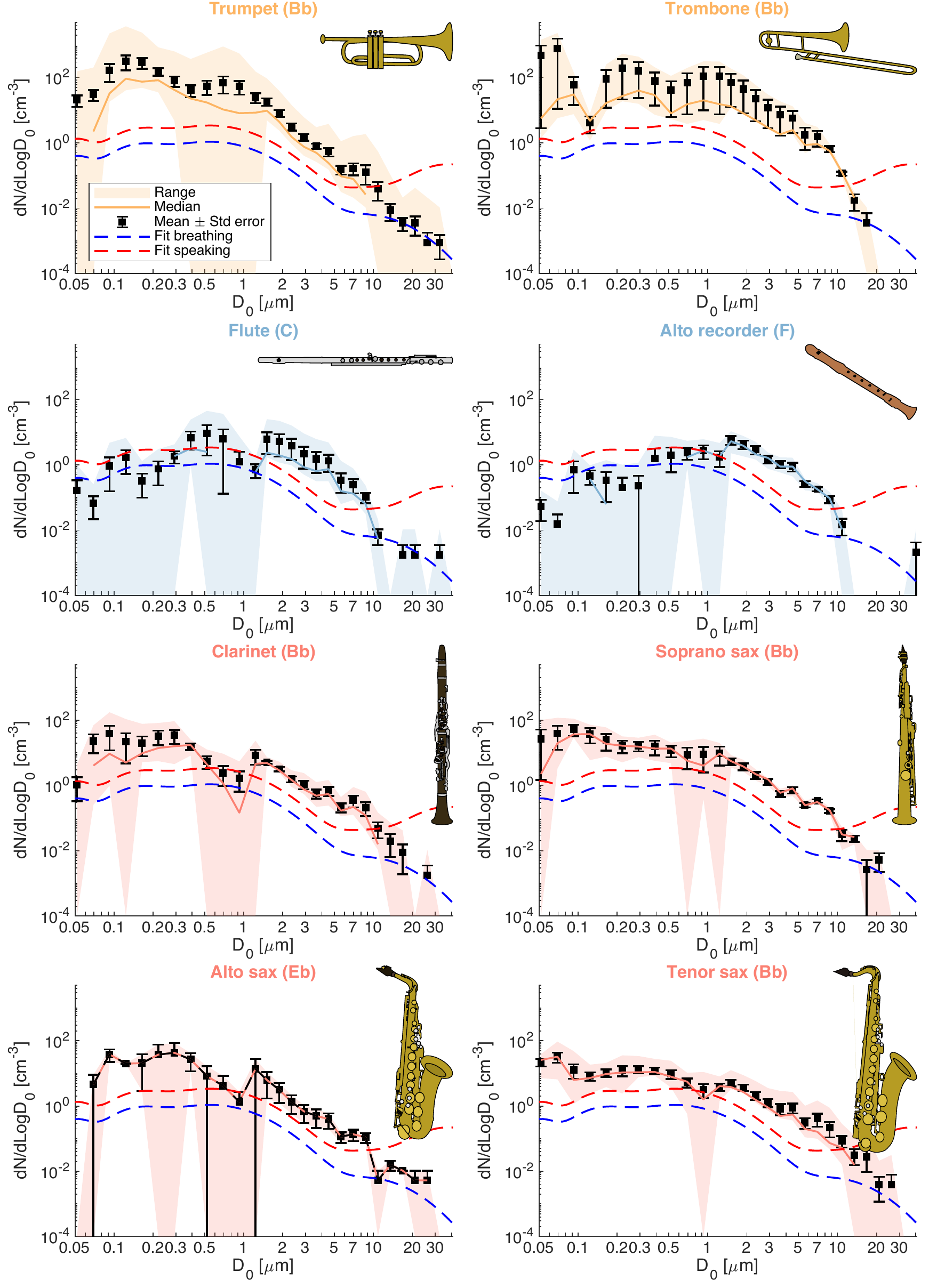} \\
	\caption{Normalized emitted particle concentration (dN/dLogD\textsubscript{0}) from SMPS and OPS obtained from two brass (orange), two edge-blown woodwind (blue) and four single-reed woodwind instruments (red) as a function of exhaled diameter $D_0$. Shown are arithmetic mean and standard error (black symbols and errorbars), median (solid lines), and range from 25th percentile to maximum (shaded area). Dashed blue (Fit breathing) and red lines (Fit speaking) represent multi-modal log-normal fits for breathing and speaking as published in \cite{Bagheri2021database}. For each wind instrument type, the base or reference note is given in parenthesis.} 
	\label{fig:sizedistro}
\end{figure}

Normalized number concentration of exhaled particles produced during playing wind instruments measured by the OPS and SMPS as a function of exhale diameter $D_0$ are shown in Fig. \ref{fig:sizedistro}.
Particles between 0.1 and \SI{0.3}{\micro\meter} represent between 1\% and 20\% of the total volume concentration, so the particle size distribution measured by the OPS, for which more data is available, covers 80 to 99\% of the emitted volume concentration. Apart from the recorder and the flute (second row of Fig. \ref{fig:sizedistro}), the emitted concentration of particles smaller than \SI{1}{\micro\meter} while playing a wind instrument is about an order of magnitude higher compared to normal breathing found by  \citet{Bagheri2021database}. In addition, the fitted particle emission for a person speaking is shown in Fig. \ref{fig:sizedistro} as dashed blue lines. 

While the particle concentration tends to increase after \SI{10}{\micro\meter} towards larger sizes for speaking, there is a rapid decrease of particle concentration in this size range observed in the wind instrument data. As already explained in Materials and Methods, Subsection \ref{sec:experiments}, we can trust the measured particle concentrations to about \SI{16}{\micro\meter} diameter at exhalation, which is \SI{3.5}{\micro\meter} dry diameter. As the observable decrease in particle concentration released by wind instruments starts at smaller sizes, we think the decrease is real and caused by the wind instruments themselves acting as a filter. We know from the results discussed in \cite{Bagheri2021database} that a substantial amount of large ($D_0$\textgreater \SI{10}{\micro\meter}) particles is released during phonetic activities. It is only the high sampling and detection efficiency of the holographic setup that enables reliable characterization of such large particles. We have also conducted experiments with several wind instrument being played right into the sample volume of the holographic setup. These experiments were done with four trumpets, one tenor trombone, two clarinets, two transverse flutes, one alto recorder, one soprano sax, one alto sax and two tenor saxophones. In none of the experiments, we were able to detect particles \textgreater \SI{6}{\micro\meter} at the bell end of any wind instrument with the holographic setup. These results give us confidence that we are not missing an important mode of the size distribution at larger sizes. In general, we found that wind instrument players release higher concentration of particles up to \SI{10}{\micro\meter} exhaled diameter when playing their instrument compared to breathing and speaking. 

In addition to the combined SMPS and OPS data presented in Fig. \ref{fig:sizedistro}, we examined the entire set of OPS data and performed multi-modal lognormal fits to the mean size distributions per instrument group similar to the procedure described in \cite{Bagheri2021database}. The size distributions were fitted with the function described in Eq. \ref{eq:fit} where $D_i$ is the $i$-th mode diameter at exhalation, $\sigma_i$ is the geometric standard deviation of the $i$-th mode, and $A_i$ is the mode concentration. 
\begin{equation}
    dN/dLogD_0 = \sum_{i=1}^n A_i \exp(-(\ln(D_0/D_i)/\sigma_i)^2)
    \label{eq:fit}
\end{equation}

To all measured size distributions, we could fit a mode at $D_1 = $\SI{0.56}{\micro\meter} as in \cite{Bagheri2021database}. However, the geometric standard deviation $\sigma_1$ needed to be varied for the different wind instrument types to achieve a good coefficient of determination ($r^2$). To avoid overfitting, the first step was to fit the first mode to the data with fixed $D_1$ to determine $A_1$ and $\sigma_1$. Then, the fitted first mode was subtracted from the data and another mode was fitted to the remaining data. For most instrument types, two modes could explain the measured data very well. For some instruments (practice chanter, Great Highland Bagpipes, Scottish Smallpipes, F tuba), a third mode was visible and fitted to the data remaining after mode 1 and 2 were subtracted. In the final step, all $D_i$ and $\sigma_i$ were fixed, based on the results of the previous fits, and the coefficients $A_i$ were used as free parameters for the fit to the original data. All fit parameters and coefficient of determination are shown in Table \ref{tab:fitparams}. A weighting function was used for the final fit, which is defined as $1/\sigma(dNdLogD_0)$ where $\sigma(dNdLogD_0)$ is the standard error of the data. This choice was taken to yield better agreement between the fit and the measured concentration of particles \textgreater \SI{10}{\micro\meter}. 

\begin{table}[!htbp]
\caption{Parameters of the multi-modal lognormal fits to the data for each instrument type. The mode diameter $D_1$ was fixed at \SI{0.56}{\micro\meter}. If a third mode was fitted as well, the parameters of the third mode are given in the columns of $A_2$, $D_2$ and $\sigma_2$ in parenthesis.}
\begin{tabular}{ccccccc}
\hline
Instrument & $A_1$ [\SI{}{\per\cubic\centi\meter}] & $\sigma_1$ & $A_2$ [\SI{}{\per\cubic\centi\meter}] & $D_2$ [\SI{}{\micro\meter}] & $\sigma_2$ & $r^2$  \\
\hline 
Piccolo & 1.42 & 1.03 & 0.04 & 2.6 & 0.66 & 0.96\\
Fife & 1.92 & 1.15 & 0.05 & 3.8 & 0.55 & 0.96\\
Flute & 8.6 & 0.96 & 0.58 & 2.7 & 0.75 & 0.97\\
Recorder & 20.1 & 0.89 & 0.94 & 2.7 & 0.74 & 0.98\\
\hline
Clarinet & 44.1 & 1.05 & 1.42 & 5.9 & 0.58 & 0.99\\
Sop. sax &14.5 & 1.03 & 0.47 & 5.1 & 0.71 & 0.99\\
Alt. sax &12.3 & 0.93 & 0.07 & 2.8 & 1.27 & 0.97 \\
Ten. sax &13.3 & 0.90 & 0.43 & 3.3 & 0.91 & 0.99 \\
\hline
Pr. chanter & 484 & 0.82 & 2.31 (0.004) & 2.9 (20.1) & 0.54 (0.41) & 0.98\\
Bagpipes & 24.5 & 1.11 & 0.05 (0.006) & 3.0 (27.0) & 0.59 (0.40) & 0.99\\
Smallpipes &1.86 & 1.10 & 0.08 (0.009) & 3.5 (19.5) &0.81 (0.48) & 0.98\\
\hline
P. trumpet & 30.0 & 0.97 & 0.27 & 3.5 & 1.16 & 0.98\\
Trumpet & 75.5 & 0.84 & 0.31 & 3.9 & 1.01 & 0.99\\
Kuhlo horn & 27.0 & 1.02 & 0.78 & 5.9 & 0.66 & 0.99\\
Hunting horn & 22.2 & 0.89 & 0.02 & 6.4 & 1.18 & 0.94\\
Trombone & 143 & 0.95 & 1.34 & 5.0 &0.63 & 0.99 \\
Tenorhorn & 83.4 & 0.80 & 0.62 & 1.8 & 0.91 & 0.99\\
Double horn & 10.4 & 1.01 & 0.29 & 6.8 & 0.74 & 0.99\\
F tuba & 49.5 & 0.90 & 1.96 (0.004) & 2.8 (30.8) & 0.75 (0.30) & 0.99\\
Bb tuba & 5.5 & 1.06 & 0.06 & 3.8 & 0.67 & 0.99\\
\hline
\end{tabular}
\label{tab:fitparams}
\end{table}

\begin{figure}[htbp]
	\centering
	\includegraphics[width=1.0\textwidth]{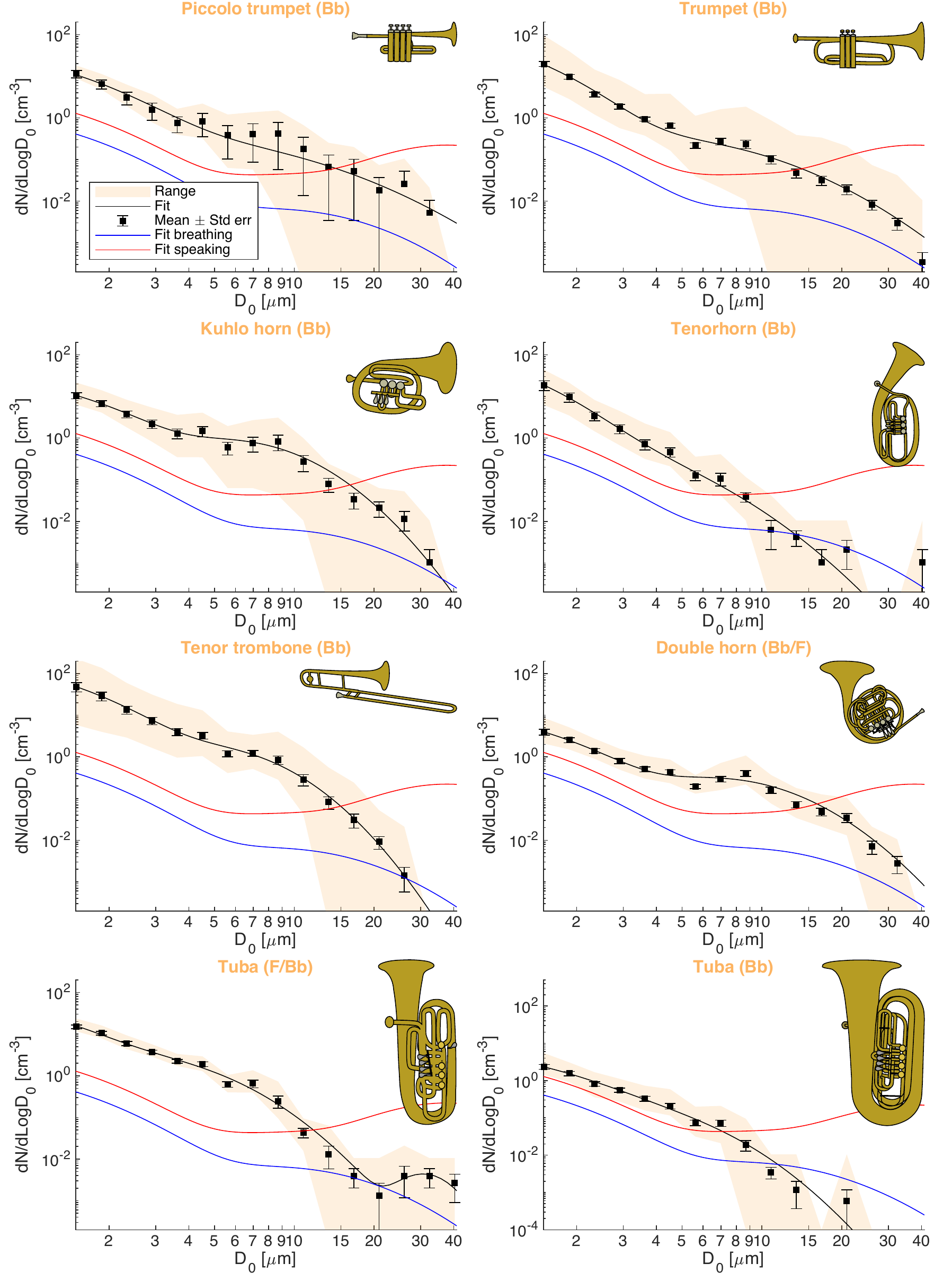} \\
	\caption{As in Fig. \ref{fig:sizedistro} for eight brass instruments in the measurement range of the OPS. Mean values and standard error are shown by the black symbols with errorbars, the black solid line represents a multi-modal lognormal fit to the data.} 
	\label{fig:psdbrass}
\end{figure}

\begin{figure}[htbp]
	\centering
	\includegraphics[width=1.0\textwidth]{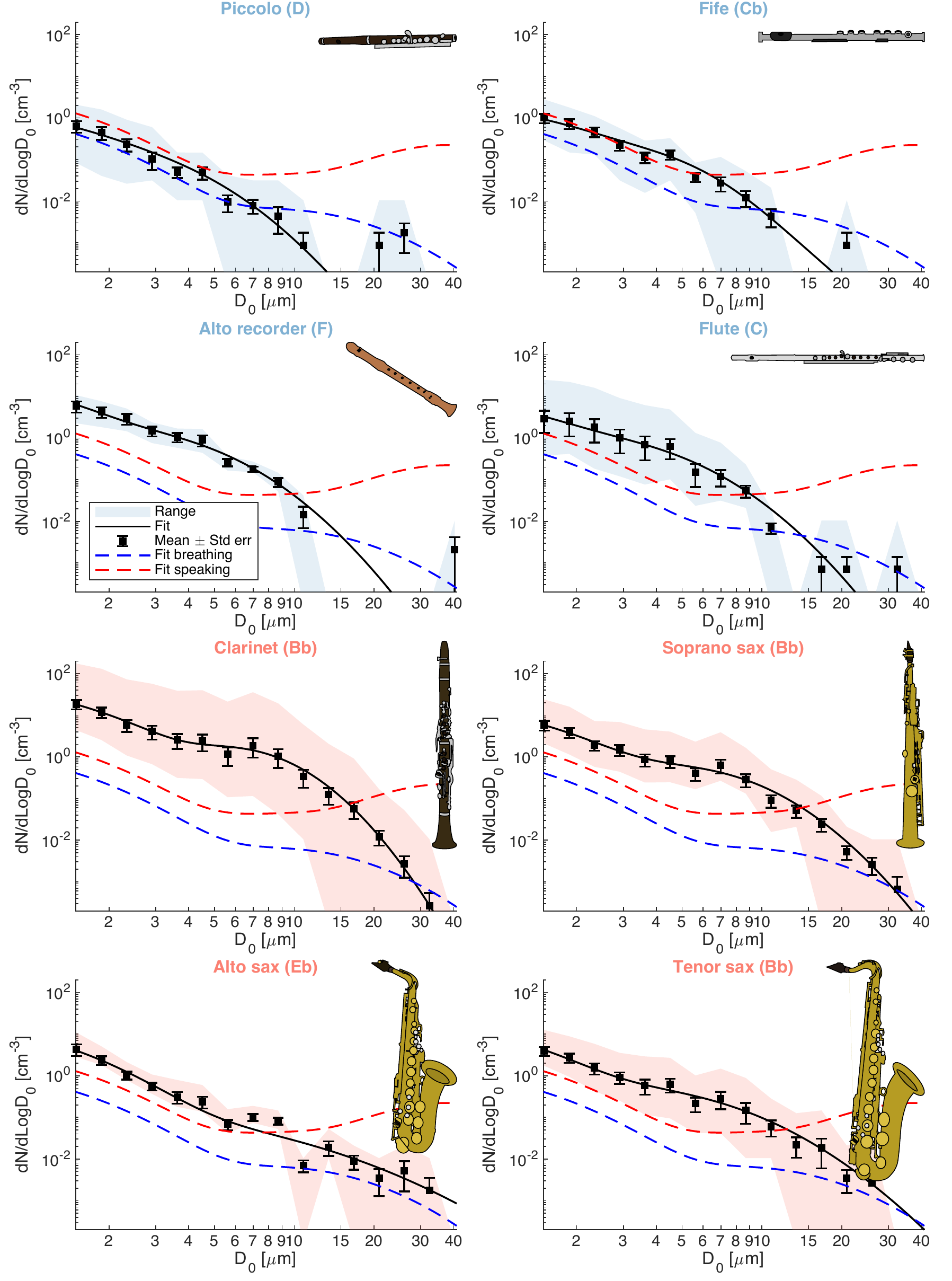} \\
	\caption{As in Fig. \ref{fig:psdbrass} for eight woodwind instruments (edge-blown woodwinds in blue, single-reed woodwinds in red).} 
	\label{fig:psdwoodwind}
\end{figure}

\begin{figure}[htbp]
	\centering
	\includegraphics[width=1.0\textwidth]{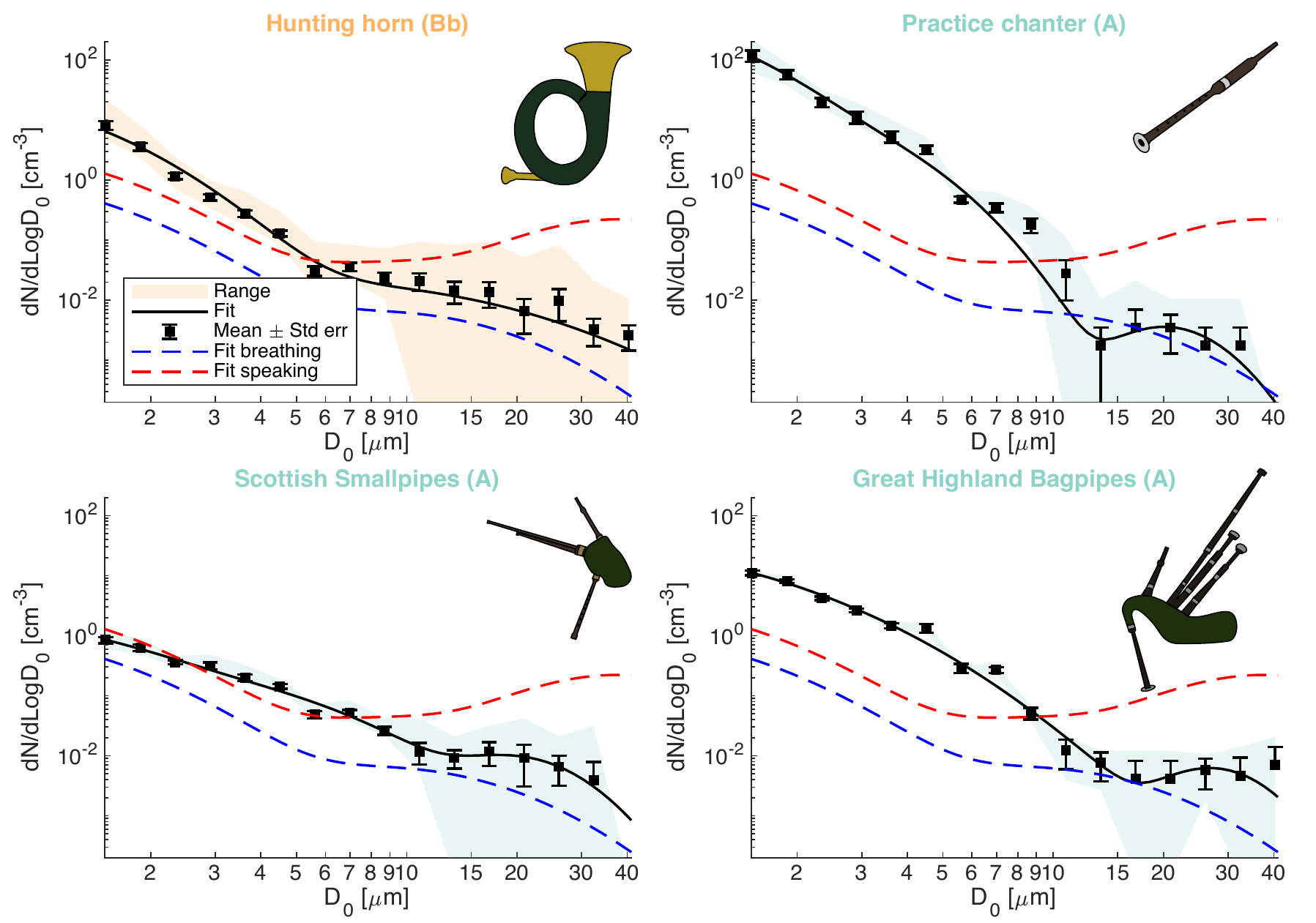} \\
	\caption{As in Fig. \ref{fig:psdbrass} for four other instruments including hunting horn (brass, in orange), practice chanter (directly mouth-blown double-reed, in green) and two bagpipes (indirectly mouth-blown double-reed, in green).} 
	\label{fig:psdother}
\end{figure}

It needs to be emphasized that the data presented in Fig. \ref{fig:sizedistro} are those data where the SMPS provided valid size distributions. Size distributions measured by the OPS, including subjects and instruments without valid SMPS data, are shown in Fig. \ref{fig:psdbrass}, \ref{fig:psdwoodwind} and \ref{fig:psdother}. The general trend of most wind instrument players emitting much higher particle concentration when playing the instrument compared to normal breathing is also visible in Fig. \ref{fig:psdbrass} - \ref{fig:psdother}. 

We found that playing edge-blown woodwind instruments tends to produce less particle number-concentration compared to playing brass instruments or reed instruments (see Fig. \ref{fig:psdwoodwind}). Both piccolo and fife had emitted size distributions very similar to the size distribution emitted during normal breathing in terms of shape. In direct comparison between wind instruments in Fig. \ref{fig:psdbrass} - \ref{fig:psdother} and phonetic activities in Fig. \ref{fig:psdother}, most brass and reed instruments emit a higher number of particles per volume in the size range from \SI{1.5}{} to \SI{10}{\micro\meter} than a person speaking or singing (the total volume emitted, which is most important for the risk of infection, is discussed later). There does not appear to be a direct correlation between the instrument-tube length and the concentration of emitted particles. 

From Fig. \ref{fig:psdbrass} it can be seen that the highest number-concentration is for the tenor trombone, which has about double the tube length of a trumpet or Kuhlo horn, and four times the tube length of a piccolo trumpet. The emitted particle concentration from a F-tuba is not so much different from the particle emission by a Kuhlo horn or a trumpet despite the tube of the tuba being three times as long. In the group of the reed instruments in Fig. \ref{fig:psdwoodwind} (clarinet and saxophones) we find the highest variability in particle emission in the clarinet data. The mean emitted particle concentration from the soprano sax is somewhat lower, and the alto sax and tenor sax emit less particles than the soprano sax. For the emission by the clarinet, we found a very strong pitch dependence with lowest emission in the chalumeau register and substantially higher emission in the clarion and altissimo register (see Subsection 2.11. of SI). The highest average emission was found for the tenor trombone and the clarinet. 
However, we found that a large amount of particles is produced primarily by the vibrating lips when playing a brass mouthpiece. The effect of the instrument resonator on the emitted particle size distribution is discussed in Subsection \ref{sec:origin}. The fact that we could not detect particles with the holographic setup emitted by wind instruments is in accordance with the results by \cite{McCarthy2021} who used water-sensitive cards and also could not detect particles \textgreater \SI{20}{\micro\meter} containing water at the bell end of any wind instrument.

\subsection{Parameters influencing particle emission and significance of leakage}

Similar to the particle emission from breathing and phonetic activities as shown in \cite{Bagheri2021database}, we found that the emission measured from a given wind instrument also depends on the player. If a person has for example an emitted particle concentration for normal breathing which is higher than the geometric mean plus the geometric standard deviation of emitted particle concentration in the examined cohort for the same activity, this person is called a super-emitter \cite[see][]{AsadiEtAl2019, Bagheri2021database}. Our data indicate that this is also the case here, i.e. some musicians produce more respiratory particles with the same instrument and playing technique as others, just because of different physiological properties (see Subsection 2.9. of SI for more detailed discussion). In addition to physiological reasons for higher or lower emission, the music that was played does also have an influence on the particle emission. We typically found higher particle emission in the high register for brass players as well as for clarinet players (see Fig. 9 of SI). For other wind instruments, the results were inconclusive (details in Subsection 2.11. of SI). 

\begin{figure}[htbp]
	\centering
	\includegraphics[width=1.0\textwidth]{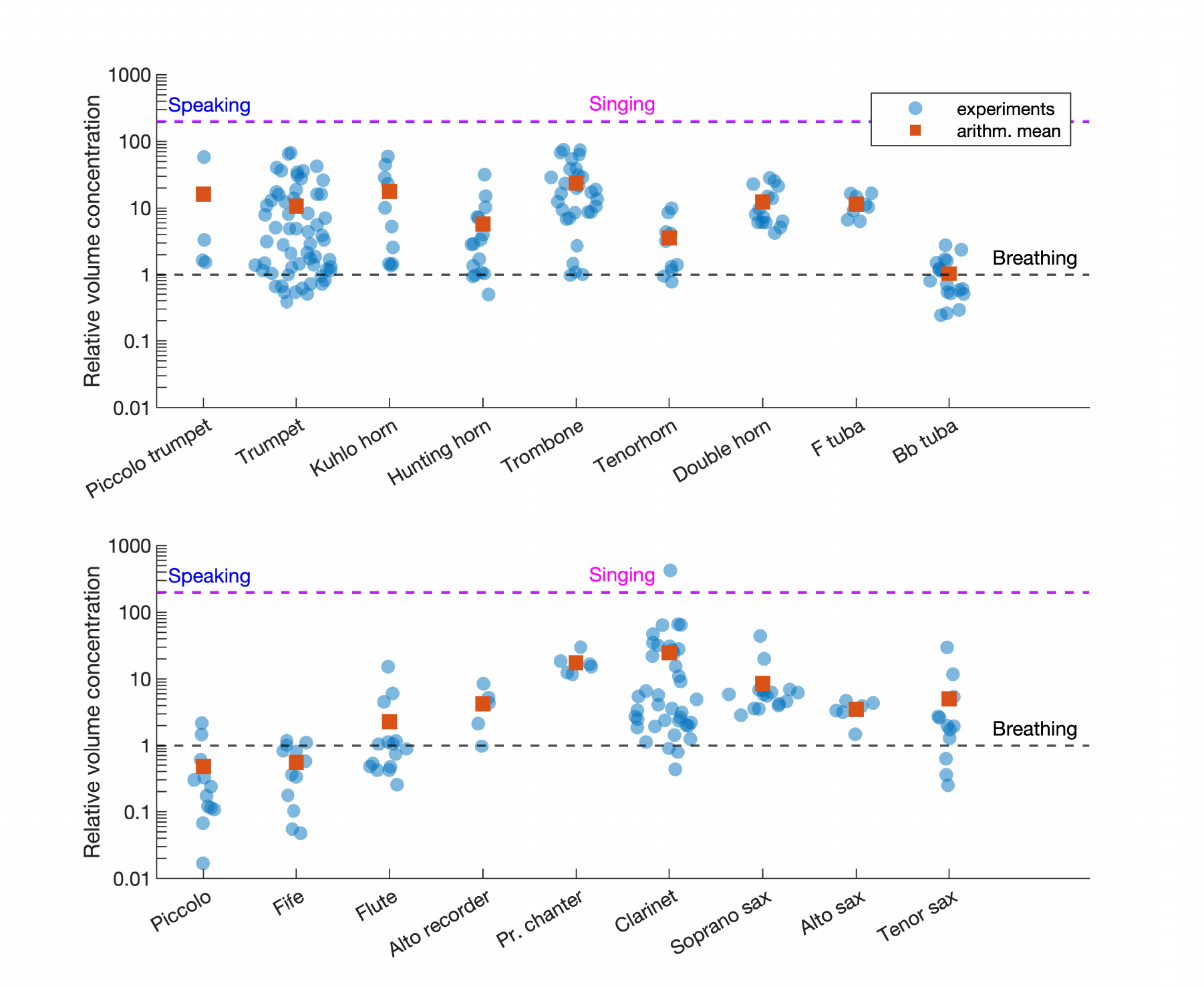} \\
	\caption{Relative emitted particle volume concentration from directly mouth-blown wind instruments (blue circles, brass in top panel, woodwind in bottom panel) compared to normal breathing (dashed black lines), speaking (dashed blue lines) and singing (dashed magenta lines). Each blue circle represents one experiment (which is one subject-instrument-music combination at some time). Red squares denote the arithmetic mean for the respective wind instrument. "Pr. chanter" is the abbreviation for "practice chanter". Relative volume concentrations for singing and speaking values are from \cite{Bagheri2021database} for particles with \SI{1.3}{\micro\meter}$<D_0<$\SI{45}{\micro\meter}, which are almost indistinguishable from each other in this plot.} 
	\label{fig:violin}
\end{figure}

To show the combined variability for most of the instruments, we present the total emitted volume concentration of particles \SI{1.3}{\micro\meter}$<D_0<$\SI{45}{\micro\meter} for all subject-instrument-music combinations of playing the wind instrument normalized to that of breathing (the mean of more than 130 subjects aged between 5 and 80 years) in Fig. \ref{fig:violin}. Compared with the dashed black line, which marks a relative emitted volume concentration of 1 indicating the same volume concentration to that of breathing, we see that most wind instruments typically are above. However, the mean volume concentration produced by wind instruments is well below that of the speaking or singing (if we consider only particles with $D_0<$\SI{45}{\micro\meter}). Only the piccolo and the fife have an emitted volume concentration that is mostly below the emitted volume concentration from normal breathing. In case of the flute, the data appear to be clustered with some data points at rather low emitted volume concentration and some at high emitted volume concentration. Here we need to mention that two of the flutists were super-emitters in normal breathing but only one of them had a comparably high emission when playing the flute. The between-subject variability of emitted volume concentration in the cohort of the flutists for playing the same note on their flutes was as high as a factor of 20 between the lowest and the highest emitter. On the other hand, we saw that two super-emitters in normal breathing in the cohort of trumpet players were not the subjects with the highest emission when playing the instrument. This was also the case for the clarinet, where the one super-emitter in normal breathing had much lower emission when playing the clarinet compared to the subject who had the highest emission recorded during playing clarinet. 

Between the different classes of wind instruments, we saw some trends. For example, the emission measured from edge-blown woodwind instruments (piccolo, fife, flute, recorder) was on average lower compared to most brass instruments (except for the B-flat tuba) and single-reed woodwind instruments. A possible explanation is the generation of the sound for these different instrument types. Edge-blown woodwind instruments need a fast, narrow airstream with virtually no resistance to make a good sound whereas the lips of a brass player or the mouthpiece of a single-reed woodwind instrument have a notable resistance to the exhaled air. The pressure within the respiratory tract needs to be higher for playing a brass or a single-reed woodwind instrument compared to playing an edge-blown woodwind instrument and we already found an increase in emitted particle concentration for exhalation at higher pressure (see Subsection 2.7 of SI).

\subsection{Particle origin for different classes of wind instruments}
\label{sec:origin}

\begin{figure}[htbp]
	\centering
	\includegraphics[width=1.0\textwidth]{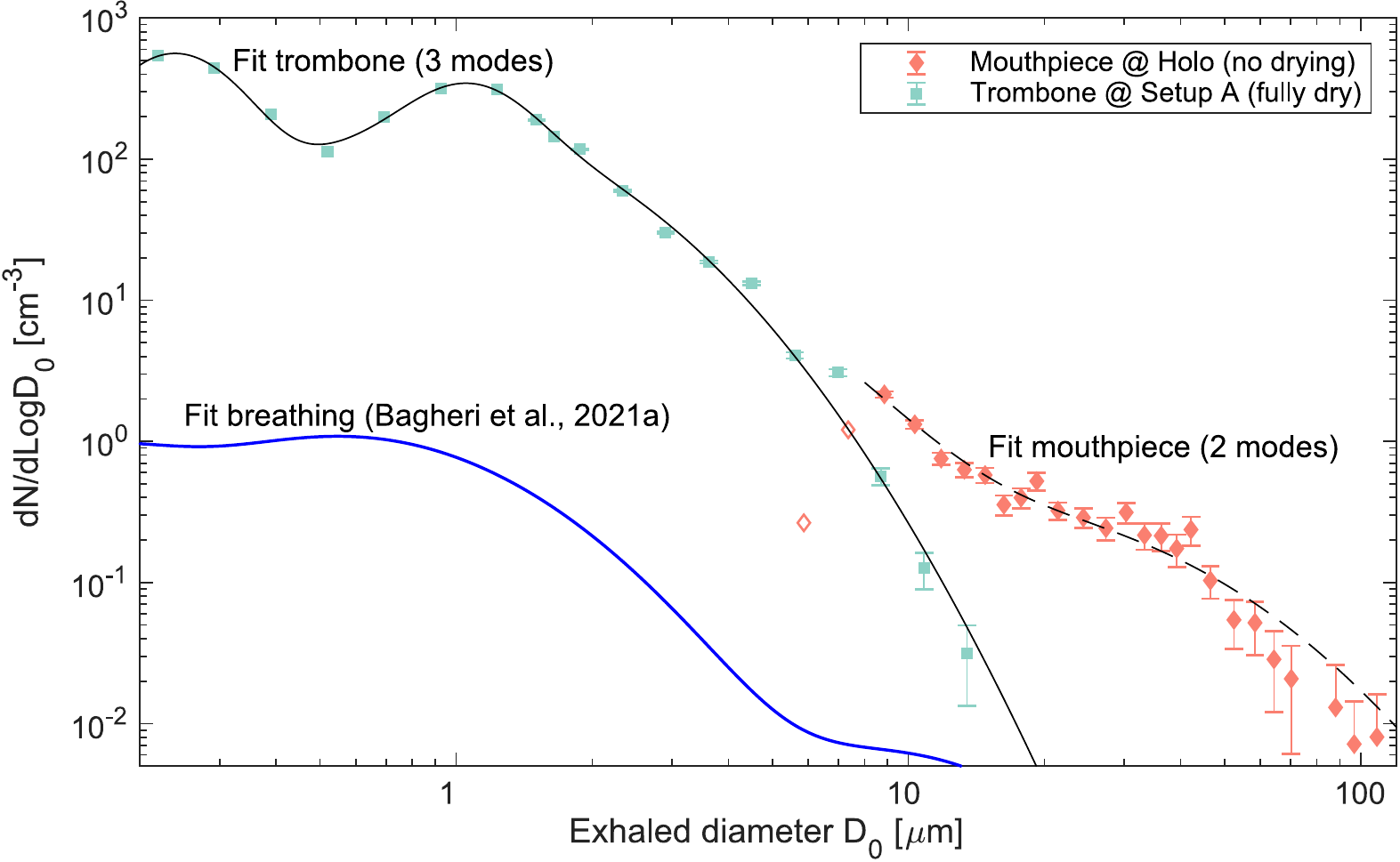} \\
	\caption{Particle size distributions measured from a trombone mouthpiece directly with the holography setup (red, uncorrected due to no drying). For comparison, the exhaled particle size distribution measured at the trombone bell via the standard setup (green, fully dry and corrected) is shown. The two first unfilled data points of the holography setup indicate the size range with poor detectability. Errorbars show the variability from counting statistics. Multi-modal log-normal fits to the bell end data (solid black line) and mouthpiece data (dashed black line) are shown in comparison with fitted size distribution for breathing (blue) in \cite{Bagheri2021database}.} 
	\label{fig:origin}
\end{figure}

The fact that edge-blown instruments typically release less particles compared to reed instruments and brass instruments when being played can be explained by the air pressure that is needed to generate the sound. On the other hand, we know from \cite{Bagheri2021database} that the oral cavity and the lips are very important sources of particles released by a person speaking or singing. When playing a brass instrument, the sound is generated by the vibrating lips. If they are wetted before they are opened and closed, they can produce a large amount of particles \textgreater \SI{30}{\micro\meter}. We investigated the production of particles by a brass player with the holographic setup and OPS-based measurements of particles that passed the mouthpiece of a trombone. To study the role of the resonator (the trombone), we also measured the particle size distribution at the trombone bell end and recalculated the particle size at exhalation from the measured dry diameter as shown in Fig. \ref{fig:origin}. The size distributions are based on three subsequent experiments executed by the same subject playing the same piece of music for the same duration at same dynamics, which were (i)\enquote{Mouthpiece @ Holo}: playing the trombone mouthpiece into the holographic sample volume (no correction needed due to negligible drying), (ii)\enquote{Mouthpiece @ OPS}: playing the trombone mouthpiece into the OPS inlet (cannot be corrected due to partial drying), and (iii)\enquote{Trombone @ Setup A}: playing the trombone into the standard setup (corrected assuming full drying to obtain $D_0$). 
As expected from the size distributions in Fig. \ref{fig:sizedistro}, the particle concentration emitted by the trombone player through the instrument is about two orders of magnitude higher than the typical emission from breathing. Around \SI{6}{\micro\meter} exhaled diameter, the fitted data from the trombone mouthpiece measured by the holographic setup intersect with the corrected (measured dry) data from the trombone bell end measured via Setup A. Due to partial drying, it was not possible to correct the data measured by the OPS directly from the trombone mouthpiece. What we can derive from the data shown in Fig. \ref{fig:origin} is that the trombone acts as a filter with a transmission $T$ of $\sim 1$\% for $D_0$ \textgreater \SI{20}{\micro\meter} and $\sim 0.1$\% for $D_0$ \textgreater \SI{30}{\micro\meter}. It should be emphasized that up to \SI{20}{\micro\meter} exhaled diameter, the data obtained via Setup A are trustworthy. Moreover, it is actually possible to approximate the transmission of the trombone resonator by $T \sim \exp(-0.3 (D_0 - D_1))$ with $D_1 = $ \SI{3.8}{\micro\meter}. This is the same functional relationship as found for penetration of non-charged melt-blown filter fabrics, the only difference is that $D_1 = 0$ for the filter fabrics (see Subsection \ref{sec:masks} and Subsection 2.13 of SI for more details). 

However, if these particles are captured somewhere in the resonator, they might eventually become resuspended. Also, the removal of large particles by the resonator (most likely via wet deposition) can explain the minimal difference in mean emitted particle volume concentration between reed and brass instruments. We think that the particle production mechanism for \textgreater \SI{5}{\micro\meter} exhaled diameter is the same in both brass and reed instruments. While the lips, which are usually covered with some saliva, open and close up to several hundred times per second when playing a brass instrument, the same happens with the mouthpiece of a single-reed instrument, or with a double-reed if the reed is covered with some saliva.

\FloatBarrier
\subsubsection{Leakage}

We found that concentration of particles leaking from the tone holes of woodwind instruments is in some cases about as much as the that of the bell end, and they have a very much comparable particle size distribution (see Fig. 8 and Subsection 2.5. of SI). Brass instruments do not have leakage at the valves, but another source of infectious particles is particles released by normal breathing, either through the nose or the mouth. The particle leakage from tone holes is taken into account in our infection risk calculations by assuming the upper bound of exhalation volume flow rate for professional wind instrument players given by \cite{Bouhuys1964} for different instrument types. For some of the brass and woodwind instrument players, we examined the size distribution of particles from mouth or nose during the play with an additional measurement setup and found that the particle size distributions were very close to the particle size distributions from normal breathing (see Fig. 7 and Subsection 2.5. of SI). From these findings, two consequences arise: Firstly, woodwind instruments need to be covered entirely with an instrument-mask as leakage from the tone holes is an issue. Secondly, particle emission from normal breathing while playing the instrument needs to be taken into account for the calculation of infection risk. 

\subsection{Masks for wind instruments}
\label{sec:masks}
\begin{figure}[htbp]
	\centering
	\includegraphics[width=1.0\textwidth]{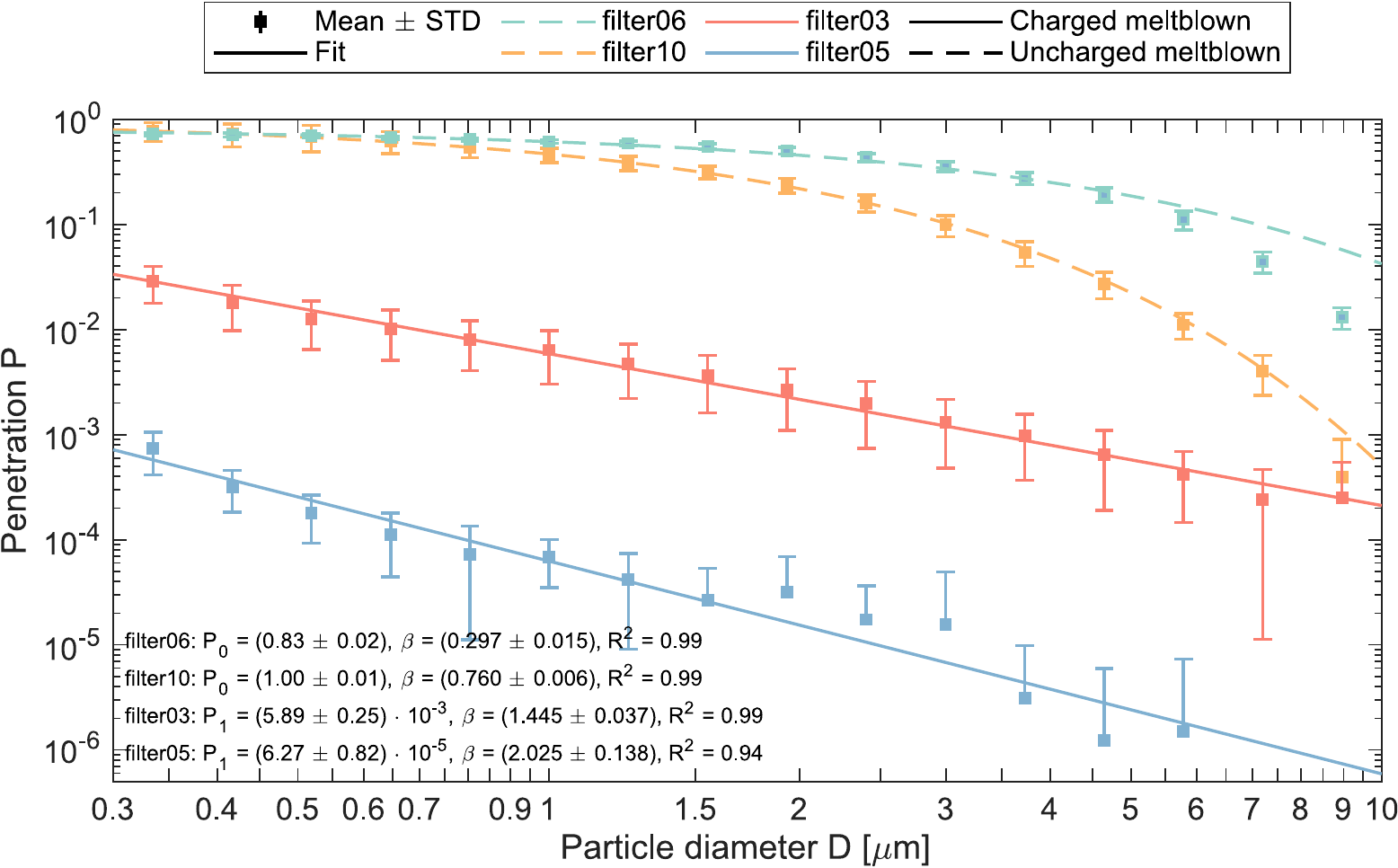}
	\caption{Measured penetration for two uncharged fabrics (filter06, filter10), and two charged fabrics (filter03 and filter05). Shown are mean penetration (squares), standard deviation (errorbars) and fitted penetration with parameters given in the lower left corner (solid lines).} 
	\label{fig:filter}
\end{figure}

We have also investigated the possibility of using instrument-masks to reduce dispersion of potentially infectious particles. Different melt-blown fabrics were tested on various wind instruments and their size-dependent penetration was measured with dry dolomite dust. In Fig. \ref{fig:filter} we show four examples of the tested fabrics. As visible in the dashed versus solid lines, charged melt-blown and uncharged melt-blown fabrics have a different functional relationship between penetration and particle size. Penetration $P$ as a function of particle diameter $D$ follows either a power law as expressed by Eq. \ref{filtration_charged} if the fabric is charged, or, in case of an uncharged fabric, an exponential decay as expressed in Eq. \ref{filtration_uncharged}. $P_0$, $P_1$ and $\beta$ are the fabric-specific parameters. More details can be found in Subsection 2.13 of SI. 

\begin{align}
P(D) = & P_1 \cdot (D/\SI{1}{\micro\meter})^{-\beta} \label{filtration_charged} \\
P(D) = & P_0 \cdot \exp \left({-\beta} \cdot (D/\SI{1}{\micro\meter}) \right)
\label{filtration_uncharged}
\end{align}

Due to wind instruments emitting the largest fraction of volume concentration in a size range $D_0$\textless \SI{10}{\micro\meter} (i.e. \textless \SI{2.2}{\micro\meter} fully dried diameter), it is utmost important to use charged melt-blown fabric for instrument-masks. The acoustic properties as described by the musicians can be summarized in fabrics with higher weight per area yielding higher playing resistance and stronger acoustic attenuation. The mask materials had a typical pressure drop between 7 and \SI{45}{\pascal} for a mean air velocity of \SI{7.3}{\centi\meter\per\second} through the mask. This velocity corresponds to about \SI{100}{\liter\per\minute} of breathing through a typical FFP2 mask. However, the material with lowest measured penetration was quite thick (about \SI{120}{\gram\per\square\meter}) but the sound quality was acceptable for low brass instruments (trombones, tenorhorns, tubas). High brass instruments (in particular trumpets) had a notable loss of brightness in the sound due to attenuation of the high harmonics. For them, a thinner material with \SI{55}{\gram\per\square\meter} (filter03) was better suited. 

\FloatBarrier
\subsection{Risk of airborne disease transmission}

With the measured emitted particle size distribution and concentration for different wind instruments and the models introduced in Materials and Methods, Subsection \ref{sec:infectionrisk}, we now calculate the risk of infection transmission for different exposure scenarios.
As we find the range of emitted volume concentration for some of the instruments to be almost a factor 1000 (e.g. clarinet) in Fig. \ref{fig:violin}, we need to use a conservative estimation of the emitted volume concentration when calculating the infection risk. The arithmetic mean values of emitted volume concentration from the trombone and the clarinet are almost equal (see the red squares in Fig. \ref{fig:violin}). These numbers will be used to represent a high emitter. As low emitter, we chose the piccolo. With these choices the typical emission from any wind instrument is within these two extremes. In scenarios with instrument-masks, we chose the \enquote{filter05} material as mask material. We assume that all inhalation rates are the same as the corresponding exhalation rates. Breathing rates are assumed to be $Q_{I,w} = $ \SI{47}{\liter\per\minute} for the infectious playing the trombone (Max), $Q_{I,w} = $ \SI{44}{\liter\per\minute} for the infectious playing the piccolo (Min), and additional $Q_{I,b} = $ \SI{9}{\liter\per\minute} for breathing. The values for the trombone and piccolo are taken from \cite{Bouhuys1964} and the value for normal breathing is taken from \cite{ICRP1994}. For the near-field exposure, we use the volume averaged concentration of pathogen copies from both sources (instrument and nose of infectious). The susceptible breathes with a constant volumetric flow rate $Q_{S,b} = $ \SI{9}{\liter\per\minute}. For the interpretation of the infection risk, it needs to be mentioned that we show the upper bound of infection risk, and that the infection risk is cumulative. An instrument lesson, band rehearsal or concert might not be the only exposure scenario during the day. So it makes sense to draw the line of an acceptable infection risk somewhere on the low end. 

\FloatBarrier
\subsubsection{Near-field exposure}

\begin{figure}[htbp]
	\centering
	\includegraphics[width=0.9\textwidth]{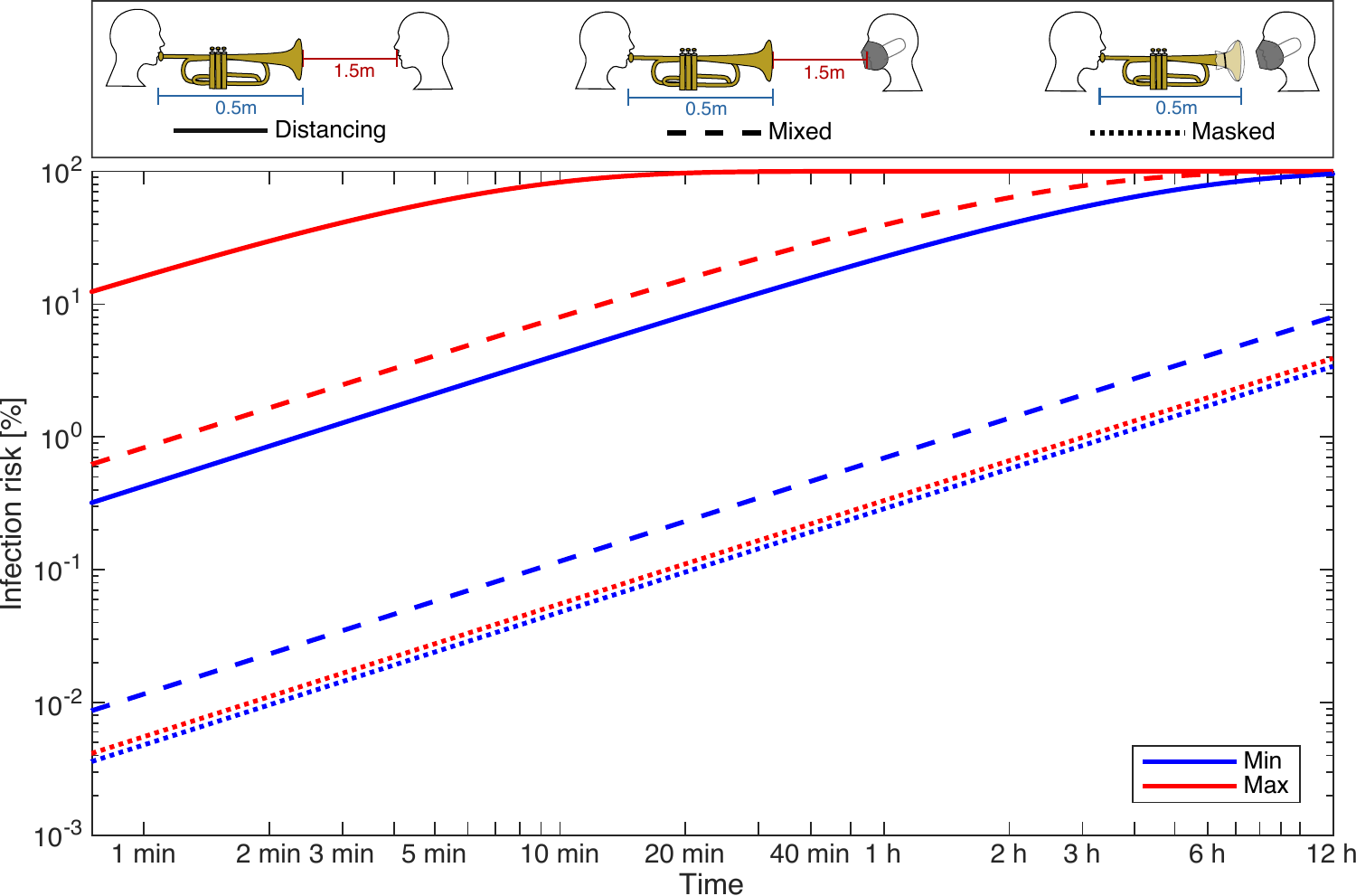}
	\caption{Infection risk for near-field exposure as a function of time. Shown are the exposure scenarios as defined in Fig. \ref{fig:scenarios} with the \enquote{Distancing} scenario (blue lines), the \enquote{Mixed} scenario (red lines) and the \enquote{Masked} scenario (black lines). Infection risk is shown for maximum (solid lines) and minimum (dashed lines) mean particle emission by the examined wind instruments; the instruments are explained in the text.} 
	\label{fig:infectionrisk1}
\end{figure}

Figure \ref{fig:infectionrisk1} shows clearly that there is already an upper bound of 50\% risk of infection transmission in the \enquote{Distancing} scenario after just 4 minutes of exposure if we consider a wind instrument with high particle emission (solid blue line in Fig. \ref{fig:infectionrisk1}). As expected, the infection risk for the same time is lower if a low emission instrument is played (dashed blue line in Fig. \ref{fig:infectionrisk1}). Though the infection risk is lower, the time to reach 50\% risk of infection is still only about 2.5 hours. 

In the \enquote{Mixed} scenario, the infection risk is largely reduced in comparison to the distancing scenario, but for a high emission instrument (solid red line in Fig. \ref{fig:infectionrisk1}), the level of protection for the susceptible is still insufficient for relatively long exposures as the infection risk is still about 50\% after 1.5 hours.

In the masked scenario, we find no difference between minimum and maximum (solid and dashed black line in Fig. \ref{fig:infectionrisk1} are on top of each other), which means that the particles emitted from mouth or nose of the infectious during breathing dominate the infection risk (see Fig. \ref{fig:infectionrisk1}). Even after an exposure time of 12 hours, the infection risk is still \textless 5\%. This is because the instrument-mask is able to reduce the particle emission from the instrument to about 1\% of that produced by breathing. 

The \enquote{leakage} from breathing could actually be reduced by the infectious musician wearing an adjusted FFP2 mask with a feedthrough for the mouthpiece, but this additional protection proved to be impractical for most settings with live music (e.g. music education, rehearsals, concerts). Nonetheless, given that we have calculated a very conservative upper bound, the Masked scenario can be assumed safe for up to several hours when the near-field transmission of pathogens is considered only.

The selection of the mask material has an influence on the risk of infection. If the thinnest of our examined materials without charge, which is \enquote{filter10}, is used for the instrument-mask, the \enquote{Mixed} scenario leads to about the same risk as the \enquote{Masked} scenario as the reduced distance increases the risk of infection by about the same amount as the mask reduces it (see Subsection 2.18 and Fig. 17 of SI). This intercomparison shows what happens if low-efficacy masks are used and no distance is kept between infectious and susceptible. Thus, it is utmost important to use charged melt-blown materials for instrument-masks. Even with a thin charged melt-blown material such as \enquote{filter03}, the \enquote{Masked} scenario is the safest of all near-field exposure scenarios.

\subsubsection{Far-field exposure}

\begin{figure}[htbp]
	\centering
    	\includegraphics[width=0.9\textwidth]{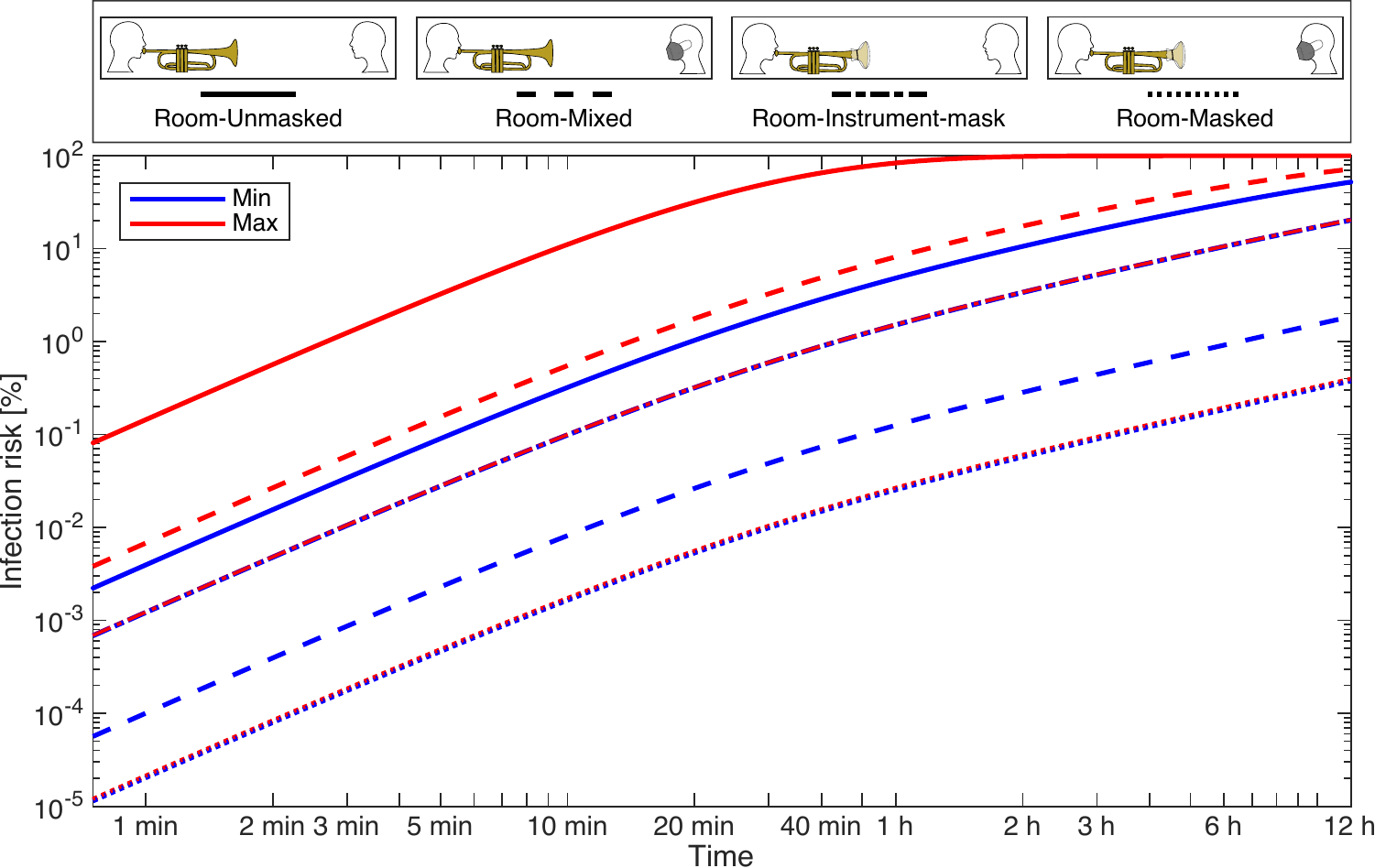}
	\caption{As in Fig. \ref{fig:infectionrisk1} for far-field exposure. Shown are \enquote{Room-Unmasked}, \enquote{Room-Mixed}, \enquote{Room-Instrument-mask} and \enquote{Room-Masked} scenario for maximum and minimum of the mean particle emission among the examined wind instruments.} 
	\label{fig:infectionrisk2}
\end{figure}

The infection risk for the far-field exposure is shown in Fig. \ref{fig:infectionrisk2}. While it took only \SI{4}{\minute} to reach 50\% infection risk in the \enquote{Distancing} scenario with the high emission instrument (see Fig. \ref{fig:infectionrisk1}), the exposure time for 50\% infection risk in the \enquote{Room-Unmasked} scenario in the far-field is about \SI{30}{\minute} (see Fig. \ref{fig:infectionrisk2}). Avoiding exposure to the near-field of an infectious increases the exposure time for the same risk by almost a factor ten, however, eliminating the risk of infection transmission associated with near-field exposure in real world conditions is not trivial to achieve. Nonetheless, the \enquote{Room-Unmasked} scenario is the one with the highest risk of infection transmission and the \enquote{Room-Masked} scenario is the safest. The \enquote{Room-Instrument-mask} scenario yields a lower infection risk after 12 hours compared to the \enquote{Room-Mixed} scenario for the high-emission wind instrument. From Fig. \ref{fig:infectionrisk2}, it can be concluded that the \enquote{Room-Masked} scenario is associated with a low risk of infection transmission ($<1\% $) even after 12 hours. 


\subsection{The HEADS web-app: investigating other far-field scenarios}
For the far-field exposure scenario, we have shown a typical example representative of some music activities in the previous section. Other scenarios, where multiple infectious and/or multiple susceptible individuals are in rooms with different ventilation characteristics, can be studied with the HEADS (Human Emission of Aerosol and Droplet Statistics) web app, which has a user-friendly interface \cite[][]{heads}. In HEADS, the effect of room size, ventilation rate, number of susceptible and infectious individuals, activities, pathogen concentration, infectious dose, pathogen deactivation rate, particle deposition, intake efficiency of the susceptible, particle drying and their re-hydration of particles in susceptible airways, and poly-pathogen aerosols on the risk of infection transmission for any airborne disease can be studied. The HEADS \enquote{Advanced View} allows also mixtures of different activities (for example, 80\% breathing, 10\% speaking, 10\% singing). Another powerful tool in HEADS is the analysis of subsequent exposure scenarios. If the relevant parameters are known, it is possible to add for example the use of public transportation to/from the concert. The HEADS web app is available in English, German and some other languages. 

\subsection{Other considerations for the infection risk assessments}

The use of conservative estimates allowed us to calculate a situation-independent upper bound of infection risk for typical SARS-CoV-2 input parameters that is reliable when the results are interpreted with statistical reasoning. However, there are fundamental limitations and uncertainties that need to be discussed. For example, in rare situations atypical values for SARS-CoV-2 can be expected to occur, e.g. $\rho_p$ for an individual with SARS-CoV-2 and considered infectious ranges from about $10^{6}$ to $10^{11}$ \SI{}{\per\cubic\centi\meter} instead of the $10^{8.5}$ \SI{}{\per\cubic\centi\meter} considered here. This span by itself introduces an uncertainty of $\pm 2.5$ orders of magnitude (factor 300) for the time to exceed a given threshold of the infection risk. In other words, a given time to exceed a threshold of the risk of infection transmission of 5 minutes could actually vary between 1 second for an infectious with extremely high virus concentration, and 1 day for an infectious with low virus concentration. Also, the value of $\textnormal{ID}_{63.21}$ is not exactly known and could vary between 100 and 1000 virions for the wild type of SARS-CoV-2 \cite[see e.g.][]{Nordsiek2020}. However, it is unlikely that the uncertainty caused by $\textnormal{ID}_{63.21}$ will affect the real-world application of the results presented, at least for the near-field scenarios, since we have already used extremely conservative estimates for other parameters.  Another aspect that introduces uncertainty to the model is that the dependence of $\rho_p$ on particle origin site and time during the course of the disease is not known. There is also a recent discussion about the inactivation rate for particles containing SARS-CoV-2, which could depend on several parameters of the room air and could be higher \cite[][]{Oswin2022}. On the other hand, simpler models like the ones used in \cite{Miller2020}, \cite{Stockman2021} and \cite{Firle2021} have the same issues while they also do not take the size to volume ratio of the particles into account. In this regard, the upper bound we have presented for the near-field calculations should remain valid as we have not considered the deactivation term.

In addition, the volumetric flow rates for exhalation are based on studies with a small number of subjects (e.g. the paper by \cite{Bouhuys1964}), however, we have used conservative estimates. Another possible limitation is the aggregation of wind instruments of the same type, e.g. the "trumpet" class consists of several different trumpet models in B-flat with both rotary and piston valves and there is evidence that the instrument geometry and/or mouthpiece geometry can have an effect on emitted particle concentration and size distribution. One trumpet player played exactly the same pieces of music on two different trumpet models at same dynamics and tempo with the result that the mean emitted particle number concentrations differed systematically by a factor 2 between the two trumpet models. The assumption of a diverging jet flow is rather crude but useful to discuss the effect of dilution as a function of distance between source and receiver. Finally, the limitations discussed need to be taken into account when interpreting the results of the infection risk calculation. Nevertheless, the infection risk calculations presented can provide very useful guidance for planning rehearsals, instrumental lessons, and live music events. 

\section{Conclusion}

We have shown that playing wind instruments releases particles in a volume concentration that is typically 0.5 to 25 times higher than the particle release from normal breathing, which is, on average, still lower than the particle release from speaking or singing. 18 out of 20 examined wind instrument types had a higher mean emitted particle volume concentration compared to breathing. Edge-blown woodwind instruments had on average lower emitted particle volume concentration compared to single-reed or double-reed woodwind instruments and brass instruments. We did not observe a systematic trend in a way that for example brass instruments with a shorter tube emitted more particles than brass instruments with a longer tube. For some instruments (most brass and also clarinets), higher emission was found for higher notes played. Reed and brass instruments also had similar mean emitted volume concentration, though the playing technique is much different. 

We have also found that playing wind instruments can produce very low concentration of $D_0>$\SI{20}{\micro\meter}, which is in stark contrast to phonetic activities like speaking or singing. More than 80\% of the particle volume emitted from wind instruments lies within a particle diameter range from \SI{0.3}{} to \SI{4.5}{\micro\meter} after fully dried, i.e. $D_0\sim$1.5 to \SI{20}{\micro\meter}. 

Different melt-blown fabrics have been examined as candidates for instrument-masks. We found that only charged melt-blown fabrics can substantially reduce emission of infectious particles. The material penetration as a function of particle size follows a power law for charged materials, and an exponential decay for uncharged materials. This explains why uncharged mask materials have such a low capture efficacy for particles \textless \SI{1}{\micro\meter}. We tested a mask material with penetration \textless $10^{-4}$ for \SI{1}{\micro\meter} particles on several wind instruments and found minor influence of the instrument-mask on the sound quality of brass instruments playing in the same tonal range as a tenor trombone or lower. A practical solution for reed instruments still needs to be found as leakage from tone holes had similar particle concentration and size distribution as emission measured at the instrument bell end. 

The emitted particles may pose a substantial risk of infection transmission in a near-field exposure as well as in the well-mixed room considered here if the musician is infectious. Without masks, the risk of infection transmission exceeds 50 \% after 4 minutes for a high-emission instrument. Using an FFP2 mask for a susceptible at \SI{1.5}{\meter} distance to the bell end of the instrument results in 50\% risk of infection transmission after 1 hour. When susceptible wears an FFP2 mask and instrument-mask is used, transmission risk in near-field is only 5\% even after 12 hours of continuous exposure. In a well-mixed room situation (i.e. far-field scenarios) like a band rehearsal (with one wind instrument) or private instrument lessons, the risk of infection transmission is \textgreater 50\% after 20 minutes but stays below 0.5\% even after 12 hours with a moderate air exchange rate (4 ACH), instrument-mask and FFP2 masks for the attendees.

With the web-app HEADS \citep{heads}, it is possible to examine different exposure scenarios with multiple infectious sources and susceptible individuals at once, and investigate how changes in distance and masks as well as other effects, such as point-of-care testing, ventilation and particle deposition, could influence the infection risk. In this paper, we limited ourselves to a few exposure scenarios representative of instrument education or band / ensemble rehearsal to characterize the risk of infection transmission in one of these events in general and evaluate possible mitigation measures. By transferring the data published in this paper also to HEADS, we enable the reader to investigate different exposure scenarios by themselves through the HEADS easy-to-use interface. 

In this work we could quantify the upper bound of infection risk and showed that universal usage of well-fitting and well-filtering instrument and individual masks can reduce the risk of infection transmission significantly while distancing alone or non-universal masking are found to be high risk measures. 

\section*{Author contributions}

OS, GB and BT wrote the draft of the main paper and the Supplementary Information. OS, GB, EB, RM, SS and LT designed the experimental procedures for the subject measurements. 
SS advised all medical aspects of the investigation. EB wrote the ethics application. OS, LT, KS, BT, JK, EB and GB conducted the experiments on human subjects. OS, KS, BT and EB did the filter penetration measurements. RM advised all musical aspects of the investigation and assisted in some of the experiments on human subjects. OS, BT and GB with the help of EB analyzed and interpreted the data. OS prepared the holography data. All authors wrote the final text. 

\section*{Conflict of Interest}
The authors declare that the research was conducted in the absence of any commercial or financial relationships that could be construed as a potential conflict of interest.

\section*{Ethics statement}

The non-invasive exhaled aerosol sampling study which led to this paper was approved by the Ethics Commission of the Max Planck Society (Submission 2020\_23). The participants provided their written informed consent to participate in this study.

\section*{Data availability}

A part of the datasets generated and analyzed for this study can be found in the HEADS web app [https://aerosol.ds.mpg.de]. Data not published within HEADS are available from the corresponding authors on reasonable request. 

\section*{Acknowledgments}

Without the great support of our volunteers, this study would not have been possible. We are grateful for all suggestions we received by some of the volunteers, which finally enabled us to discover some of the ultra-high emission scenarios we might have never seen otherwise. We like to thank the Posaunenwerk der Landeskirche Hannovers, the Hochschule für Musik, Theater und Medien Hannover, the Blasmusikverband Nordrhein-Westfalen e.V., and the Jacobi-Kantorei Göttingen for sharing the call for volunteers and encouraging their members to participate in our study. Special thanks go to the machine shop of the Max Planck Institute for Dynamic and Self-Organization, and Udo Schminke (head of construction) for making customized tools and adaptors needed for the aerosol measurements. We gratefully acknowledge the support by our High Performance Computation (HPC) work group to provide resources for analysis of the holographic data and keep the HPC cluster in a working state. 
We would also like to thank the companies producing melt-blown filter fabrics (namely Sandler AG, Filter GmbH Preussel Langenfeld, and Brigitte Hoffmann Filterprodukte KG) for providing test samples of different filter materials.

\section*{Funding}

We thank the Bundesministerium für Bildung und Forschung (BMBF) for funding within the project B-FAST (Bundesweites Netzwerk Angewandte Surveillance und Teststrategie) (01KX2021) within the NUM (Netzwerk Universit\"atsmedizin), Deutsche Forschungsgemeinschaft (project number: 469107130) and the Max Planck Society. The sponsors had no influence on the study design, data collection and analysis, decision to publish, or preparation of the manuscript.

\bibliography{references}

\end{document}